\documentclass[aps,nofootinbib,superscriptaddress,twocolumn,prl,10pt,floatfix]{revtex4-1}
\usepackage{mathrsfs}
\usepackage{graphicx}
\usepackage{enumitem}
\usepackage{amsmath,amssymb}
\usepackage{hyperref}
\usepackage{braket}
\usepackage{subfigure}
\usepackage{subfloat}
\usepackage{float}
\usepackage[usenames]{color}
\usepackage[normalem]{ulem}

\hypersetup{
    colorlinks=true,
    linkcolor=red,
    citecolor=blue,
}

\allowdisplaybreaks

\def\be{\begin{equation}}
\def\ee{\end{equation}}
\def\ba{\begin{eqnarray}}
\def\ea{\end{eqnarray}}

\definecolor{ForestGreen}{RGB}{36,179,0}
\begin{document}

\title{Hints of Nonminimally Coupled Gravity in DESI 2024 Baryon Acoustic
   Oscillation Measurements}
\author{Gen Ye}
\email{ye@lorentz.leidenuniv.nl}
\affiliation{Institute Lorentz, Leiden University, PO Box 9506, Leiden 2300 RA, The Netherlands}
\author{Matteo Martinelli}
\affiliation{INAF - Osservatorio Astronomico di Roma, via Frascati 33, 00040 Monteporzio Catone (Roma), Italy}
\affiliation{INFN - Sezione di Roma, Piazzale Aldo Moro, 2 - c/o Dipartimento di Fisica, Edificio G. Marconi, I-00185 Roma, Italy}
\author{Bin Hu}
\affiliation{School of Physics and Astronomy, Beijing Normal University, Beijing 100875, China}
\author{Alessandra Silvestri}
\affiliation{Institute Lorentz, Leiden University, PO Box 9506, Leiden 2300 RA, The Netherlands}

\begin{abstract}
The cosmic microwave background (CMB) and baryon acoustic oscillations (BAO) are two of the most robust observations in cosmology. The recent BAO measurements from the DESI collaboration have presented, for the first time, inconsistency between BAO and CMB within the standard cosmological model $\Lambda$CDM, indicating a preference for dynamical dark energy over a cosmological constant. We analyze the theoretical implication of the DESI BAO observation for dark energy and gravity employing a nonparametric reconstruction approach for both the dark energy equation of state $w_{\rm DE}(a)$ and the effective field theory coefficients. We find that the DESI data can rule out quintessence dark energy by indicating a crossing of the phantom divide at $z\lesssim1$. Furthermore, when analyzed within the broad context of Horndeski gravity which includes general relativity and many known modified gravity theories such as generalized Galileons, $f(R)$ and Brans-Dicke, our result implies that gravity should be nonminimally coupled to explain the observations, establishing the DESI result as the first hint of modified gravity. Based on these insights, we propose the \textit{thawing gravity} model to explain the nonminimal coupling and phantom crossing indicated by observation, which also fits better to DESI BAO, CMB and type Ia Supernovae data than $\Lambda$CDM.
\end{abstract}

\maketitle

%\section{Introduction and summary}\label{sec:introduction}

The cosmic microwave background (CMB) and baryon acoustic oscillations (BAO) measure with great precision the same sound horizon scale at  eras in the evolution of the  Universe that are separated by billions of years, thus offering one of the most stringent consistency tests of the cosmological model. CMB and BAO observations have remained consistent within the \textit{standard model} of cosmology, $\Lambda$CDM, in the past two decades, until recently inconsistency has aroused with the new BAO measurement from DESI DR1~\cite{DESI:2024mwx,DESI:2024uvr}, pointing to new physics beyond the standard model, especially dynamical dark energy (DE). Analyzed within $w_0w_a$CDM, i.e. a Universe described by general relativity (GR), filled with radiation, ordinary matter, cold dark matter (CDM) and a DE component with an equation of state $w_{\rm{DE}}(a)=w_0+w_a(1-a)$ (CPL)~\cite{Chevallier:2000qy,Linder:2002et}, dynamical DE is preferred over $\Lambda$CDM at $2.5\sigma$, $3.5\sigma$, $3.9\sigma$ in the joint analysis with CMB, BAO and the type Ia supernova (SNIa) data from Pantheon+~\cite{Scolnic:2021amr}, Union~\cite{Rubin:2023ovl} or DES Y5~\cite{DES:2024tys} respectively \cite{DESI:2024mwx,DESI:2024uvr}. This has prompted the exploration of theoretical models that could embed it (see e.g.~\cite{Orchard:2024bve,Chudaykin:2024gol, Alestas:2024gxe,Wang:2024dka,Notari:2024rti,Gialamas:2024lyw,Akarsu:2024eoo,Heckman:2024apk,Ramadan:2024kmn,Mukherjee:2024ryz,Giare:2024smz,Berghaus:2024kra,Yin:2024hba,Tada:2024znt,Chudaykin:2024gol,Bhattacharya:2024hep}), as well as works that question the robustness of the result \cite{Dinda:2024kjf,Cortes:2024lgw,Patel:2024odo,Liu:2024gfy,Efstathiou:2024dvn,Wang:2024rjd,Carloni:2024zpl,Colgain:2024xqj,Luongo:2024fww,Huang:2024qno,Jia:2024wix,Wang:2024pui,Shlivko:2024llw}. 

In this letter, we employ a \emph{nonparametric} approach to show that a generic implication of the DESI result is the  \textit{crossing of the phantom divide}; in fact, by directly reconstructing  the DE equation of state $w_{\rm{DE}}(a)$ from DESI+CMB+SNIa data, we show that it  crosses $-1$. This is of extreme theoretical importance because it implies that DE cannot be explained by a canonical scalar field (quintessence), which is the most widely studied DE theory. 
% Quintessence DE has $\rho_{\rm{DE}}+P_{\rm{DE}}=(1+w_{\rm{DE}})\rho_{\rm{DE}} = \dot{\phi}^2\ge0$ which prohibits phantom crossing. 
In fact, within quintessence, a crossing of the phantom divide corresponds to a negative sign in front of the kinetic term (ghost), which would lead to a Hamiltonian unbounded from below and severe problems both on the classical and quantum level~\cite{Creminelli:2008wc}.

This reveals the necessity of finding new explanations of DE with theoretically consistent treatment of both its background and perturbation dynamics. A natural and agnostic way to explore the broad theory landscape is offered by effective field theory (EFT) techniques, which allows to build an effective action out of all possible operators complying with the given set of symmetries for the system under consideration. Specific to cosmology, this is the EFTofDE~\cite{Gubitosi:2012hu,Bloomfield:2012ff}, which assumes full spatial diffeomorphism symmetry (the \textit{cosmological principle}) while time diffeomorphism invariance is broken in connection with a dynamical scalar field (DE or modified gravity, MG) sourcing cosmic acceleration. Further requiring that the resulting equations of motion are of second order, one can identify the most general EFTofDE action describing the dynamics of the background and linear cosmological perturbations, characterized by five mutually independent free functions of time, dubbed EFT functions. The EFTofDE action fully describes the background and linear cosmological dynamics of the broad class of Horndeski gravity models ~\cite{Horndeski:1974wa}, which includes general relativity (GR, corresponding to all EFT functions being null), $f(R)$ and Brans-Dicke. Any Horndeski theories can be mapped to EFTofDE actions \cite{Gleyzes:2013ooa,Bloomfield:2013efa} and vice versa \cite{Kennedy:2017sof}.

%In section~\ref{sec:eft}, 
We adopt a data-driven nonparametric reconstruction approach for the EFT functions, which directly connects observational data with the vast theory space of Horndeski gravity through a model agnostic EFT. We find that the nonminimal coupling of gravity is required to explain the DESI observation. 
% Crucially, $\Omega$ is a smoking gun of modified gravity with general relativity and quintessence having $\Omega=0$. 
Our discovery establishes the DESI result, once confirmed, as the \textit{first observational hint} of modified gravity. Based on the insights gained from the general EFT framework, we study an alternative theory of gravity, \textit{thawing gravity}, that can naturally explain the phantom crossing and nonminimal coupling indicated by the DESI observation.

%\section{Dataset}\label{sec:data}
{\bf Dataset -} We use the Monte Carlo Markov chain (MCMC) analysis to perform the reconstruction and fit models to data, with the MCMC setup detailed in the Supplemental Material \footnote{The Supplemental Material is included in the \texttt{arXiv} source files and includes references \cite{Gelman:1992zz,Bennett:2020zkv,Froustey:2020mcq,Akita:2020szl,Frusciante:2016xoj,DeFelice:2016ucp,Peirone:2017lgi,Gerardi:2019obr,Frusciante:2018vht,Peirone:2017ywi, Espejo:2018hxa,Adams:2006sv,Raveri:2019mxg,Frusciante:2018jzw,Deffayet:2010qz,deBoe:2024gpf,Ross:2014qpa,BOSS:2016wmc,eBOSS:2020yzd,KiDS:2020suj,DES:2021bvc,DES:2021vln,Kilo-DegreeSurvey:2023gfr}}. Unless otherwise specified, for all our analyses we use the following joint dataset:
\begin{itemize}
    \item \textbf{DESI}: The full BAO observation from DESI DR1~\cite{DESI:2024mwx}. 
    \item \textbf{CMB}: The CamSpec version of Planck PR4 high-$\ell$ TTTEEE \cite{Rosenberg:2022sdy} data; Planck 2018 low-$\ell$ TTEE~\cite{Planck:2019nip} data; CMB lensing of Planck PR4~\cite{Carron:2022eyg}.
    \item \textbf{SNIa} \footnote{There have been concerns that the evolution of the effective Newtonian constants in MG theories might change the luminosity of SNIa thus bias the results \cite{Garcia-Berro:1999cwy,Riazuelo:2001mg,Nesseris:2006jc,Wright:2017rsu}. However, given the stringent constraint on MG from the solar system experiments \cite{Will:2014kxa}, one expect the MG effect to be screened on small scales to be consistent. We therefore assume that the MG effect considered in this \textit{Letter} are all properly screened thus the SNIa luminosity is unaffected and its use is justified.}: Light curve observations of 1550 type Ia Supernovae (SNIa) compiled in the Pantheon+ sample~\cite{Scolnic:2021amr}.
\end{itemize}

%\section{nonparametric reconstruction of the equation of state
%DESI BAO beyond the CPL parameterization
%} \label{sec:wDE}
\begin{figure*}
    \centering
    \includegraphics[width=0.48\linewidth]{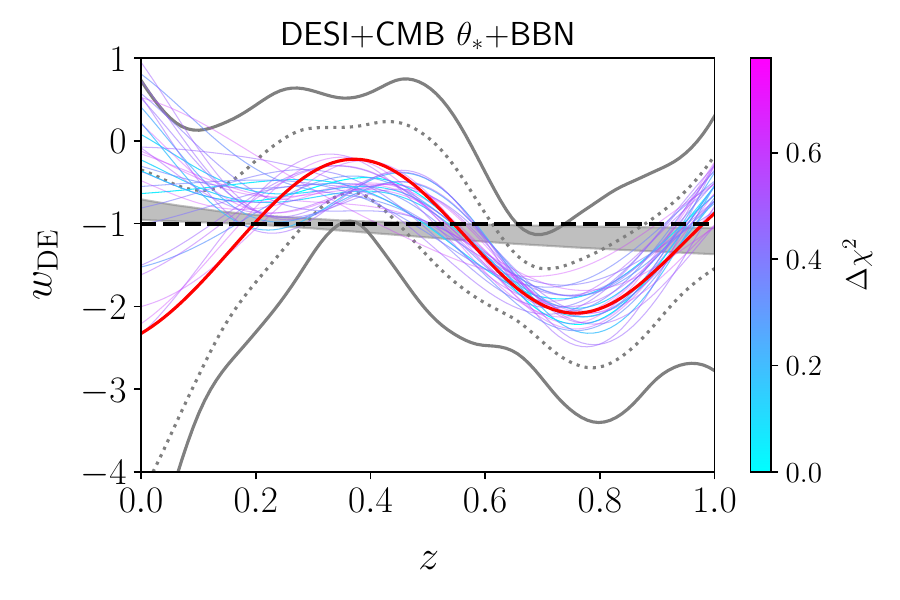}
    \includegraphics[width=0.48\linewidth]{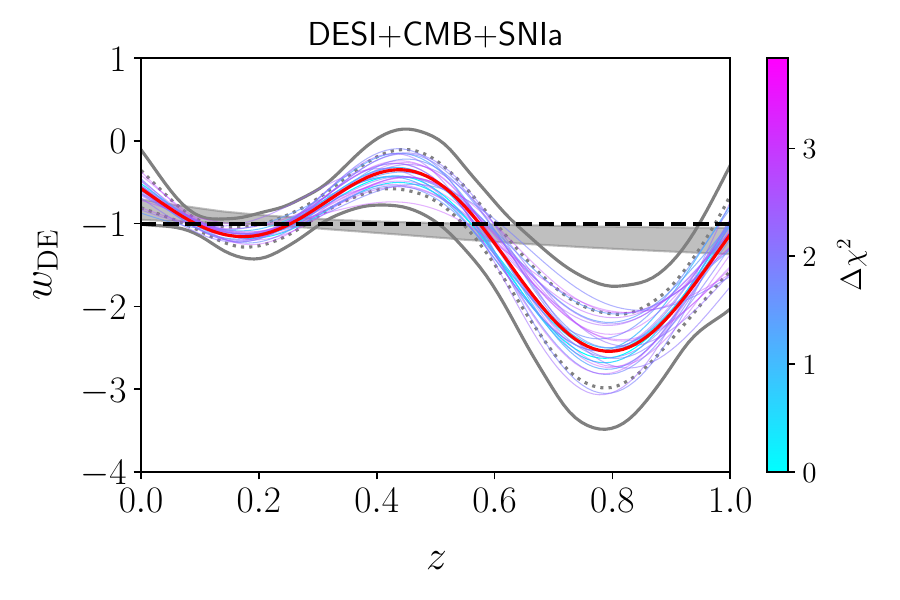}
    \caption{Reconstructions of $w_{\rm{DE}}(a)$ in the redshift range $0<z<1$. The black dashed line marks $w_{\rm{DE}}=-1$, dotted and solid gray lines indicate the 68\% and 95\% confidence interval of the reconstructed $w_{\rm{DE}}$ and the red line plots the mean function. The gray band represents the 95\% posterior region of the DESI $w_0w_a$CDM result. To further illustrate the functional shape preferred by data, we plot 50 $w_{\rm DE}$'s from the MCMC sample near the chain bestfit point in terms of $\chi^2$, with a color coding representing their $\chi^2$ difference $\Delta\chi^2=\chi^2-\chi^2_{\rm{bestfit}}$. \textit{Left panel:} Reconstruction with DESI BAO alone, aided by a BBN prior on $\omega_b$ and a CMB prior on the angular scale $\theta_*$ of recombination sound horizon. \textit{Right panel:} Reconstruction with the joint DESI+CMB+SNIa dataset.}
    \label{fig:wDE}
\end{figure*}

{\bf Nonparametric reconstruction of the equation of state -} To exploit the information in data about dynamical DE and minimize the impact of parametrization as much as possible, we reconstruct $w_{\rm DE}(a)$, without assuming any specific parametrization, by interpolation over five freely varying nodes $w_{\rm DE}(z_i)$ in the MCMC at redshifts $z_i=\{0, 0.25, 0.5, 0.75, 1\}$, roughly covering the redshift range of DESI BAO, see the Supplemental Material for detailed setup.

The reconstructed $w_{\rm{DE}}(z)$ is depicted in Fig.~\ref{fig:wDE}, with the DESI result assuming the CPL parameterization $w_{\rm{DE}}(a)=w_0+w_a(1-a)$~\cite{Chevallier:2000qy,Linder:2002et} plotted as a gray band. As our main focus here is to assess the preference for phantom crossing in the observations, rather than focusing on the detailed reconstructed shape of $w_{\rm{DE}}$, we avoid including any theory driven correlation prior in the reconstruction. This has the advantage of relying solely on information from the data but risks overfitting the data, possibly leading to the oscillatory features in Fig.~\ref{fig:wDE}, see e.g.~\cite{Raveri:2021dbu,Pogosian:2021mcs} for further discussion. Importantly, despite the small scale oscillations, the reconstructed general trend of $w_{\rm DE}$, and more specifically of its $95\%$ confidence interval, shows that $w_{\rm{DE}}>-1$ at $z=0$ and $w<-1$ for $z>0.6$  at 2$\sigma$ level, in agreement with the reconstruction from DESI using a different method \cite{DESI:2024aqx} \footnote{DESI performed a reconstruction of $w_{\rm DE}$ by expanding it in Chebyshev polynomials \cite{DESI:2024aqx}. Even though our methods differ, the results are consistent in that both imply a phantom crossing at low redshift. Ref.\cite{Yang:2024kdo} reports somewhat different results from both ours and that of \cite{DESI:2024aqx}, probably due to the different dataset used.} as well as their $w_0w_a$CDM result (gray band) \cite{DESI:2024mwx}. Assuming a continuous $w_{\rm DE}$, this leads to the parametrization-independent conclusion of phantom crossing at $z<1$.

\begin{figure*}
	\centering
	\includegraphics[width=0.48\linewidth]{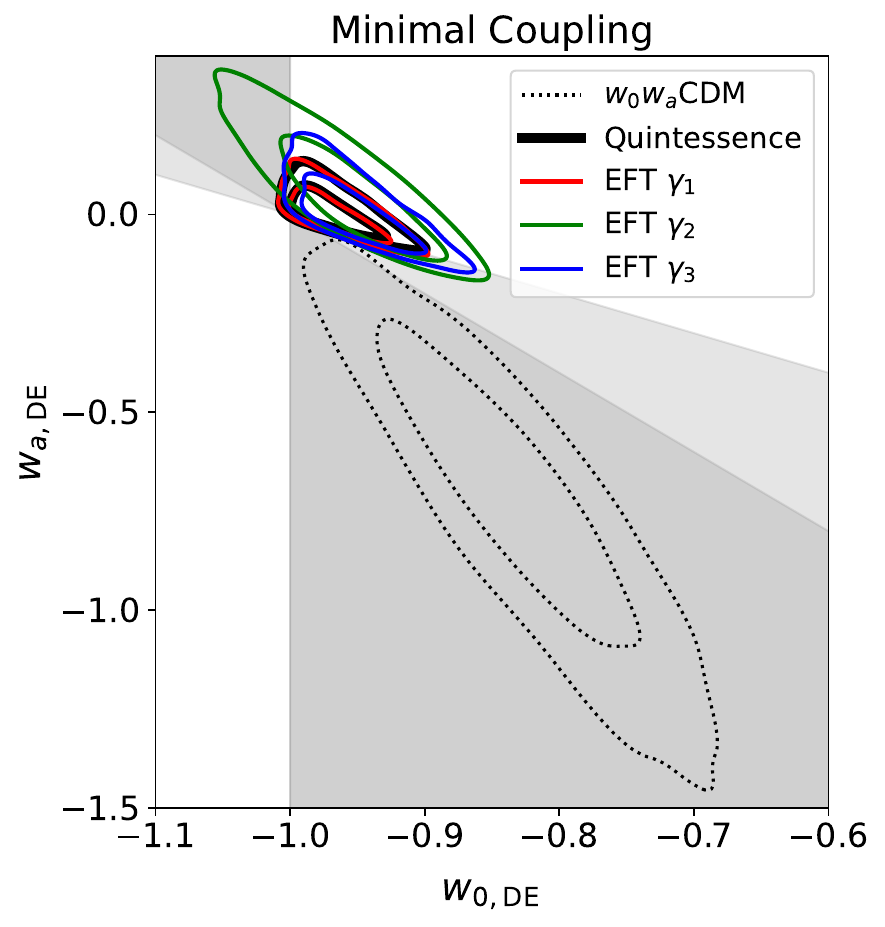}
	\includegraphics[width=0.48\linewidth]{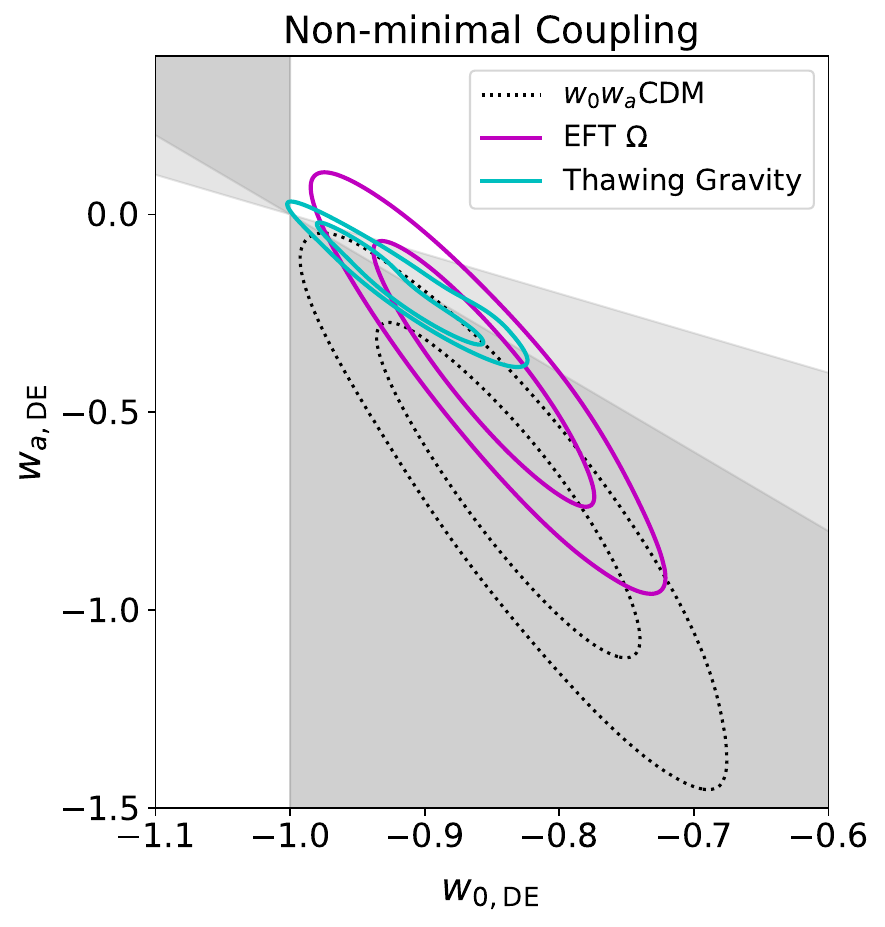}
	\caption{68\% and 95\% marginalized posterior distributions of $w_0-w_a$ over the joint dataset CMB+DESI+SNIa of all models studied in this \textit{Letter}. The reference $w_0w_a$CDM model with PPF DE perturbation is plotted with black dotted contours. The quintessence (black) contour overlaps with the EFT $\gamma_1$ result (red) so is plotted with thicker lines. Light gray shades mark the parameter regions for which phantom crossing happens, i.e. $(1+w_0)(1+w_0+w_a)<0$. Dark gray indicates regions where phantom crossing happens at low redshift,  $z<1$, i.e. $(1+w_0)(1+w_0+w_a/2)<0$ .}
	\label{fig:w0wa}
\end{figure*}

{\bf Exploration of phantom crossing within the EFTofDE - }
%\section{Exploration of phantom crossing within the EFT of DE}\label{sec:eft}
To connect with theory the observational findings, especially the phantom crossing behaviour of the effective DE fluid identified in the previous section, we survey the gravitational landscape within the general EFT of DE/MG framework, as described in the introduction. In the notation of~\cite{Hu:2014oga}, the EFTofDE has five independent EFT functions of time $\{\Lambda, \Omega, \gamma_{1,2,3}\}$. One can eliminate one EFT function, $\Lambda$, by specifying the cosmological background $H(z)$ \cite{Raveri:2017qvt}, which we do so by assuming a $w_0w_a$CDM background. Among the four remaining EFT functions $\{\Omega, \gamma_{1,2,3}\}$ to be reconstructed, $\Omega$ is the unique signature of nonminimal coupled gravity, while $\gamma_i$ signal the presence of non-standard kinetic terms and self derivative couplings of the scalar field, e.g. k-essence generally corresponds to $\Omega=0$, $\gamma_1\neq0$, $\gamma_{2,3}=0$. We reconstruct $\{\Omega,\gamma_1,\gamma_2,\gamma_3\}$ nonparametrically by binning them as functions of time using six nodes at $a\in[0.5, 0.6, 0.7, 0.8, 0.9, 1.0]$, corresponding to the redshift range $0<z<1$ most relevant to data, see the Supplemental Material for details. Despite of some concerns raised about concluding phantom crossing based on CPL~\cite{Notari:2024rti, Cortes:2024lgw, Shlivko:2024llw,Carloni:2024zpl,Wolf:2023uno}, we have shown in the previous section that the indication of phantom crossing is a property of data independent of parametrization. Therefore, we can exploit the simplicity of CPL and assume a $w_0w_a$CDM background, which also has a clearly defined phantom crossing region suitable for visualization as depicted in Fig.\ref{fig:w0wa}. We reconstruct one EFT function at a time, fixing the remaining ones to their $\Lambda$CDM value. This method allows us to isolate more clearly the function, and the corresponding EFT operator, that allows for a safe phantom crossing \footnote{We find that the constraining power of the joint dataset is not enough to reconstruct all EFT functions simultaneously, i.e. the analysis could not reach convergence within a reasonable time if more than one EFT functions are free.}.

The main result relevant to phantom crossing is summarized in Fig.~\ref{fig:w0wa}, which for reference also includes a $w_0w_a$CDM cosmology with perturbation described by the parametrized post-Friedmann (PPF) approach \cite{Hu:2007pj} and a quintessence embedding of $w_0w_a$CDM, obtained by setting the four EFT functions $\{\Omega,\gamma_1,\gamma_2,\gamma_3\}$ to their vanishing GR limit. More results are collected in the Supplemental Material. In Fig.~\ref{fig:w0wa}, the posteriors of $\{\gamma_{1,2,3}\}$ are limited by the light gray region thus these EFT functions, when turned on separately, are unable to reliably stabilize the phantom crossing indicated by observation. More importantly, our results identifies $\Omega$ as the \textit{only} operator able to lead the sampled EFT into the dark gray phantom crossing region where the DESI result resides. As $\Omega$ is the \textit{only} EFT operator related to the nonminimal coupling of gravity, we therefore arrive at the enticing conclusion that \textit{gravity should be modified to explain the recent observations}, establishing the DESI result as the first observational hint of modified gravity.

{\bf Thawing gravity - }
%\section{Thawing gravity}\label{sec:model-cov}
\begin{figure*}
    \centering
    \includegraphics[width=0.43\linewidth]{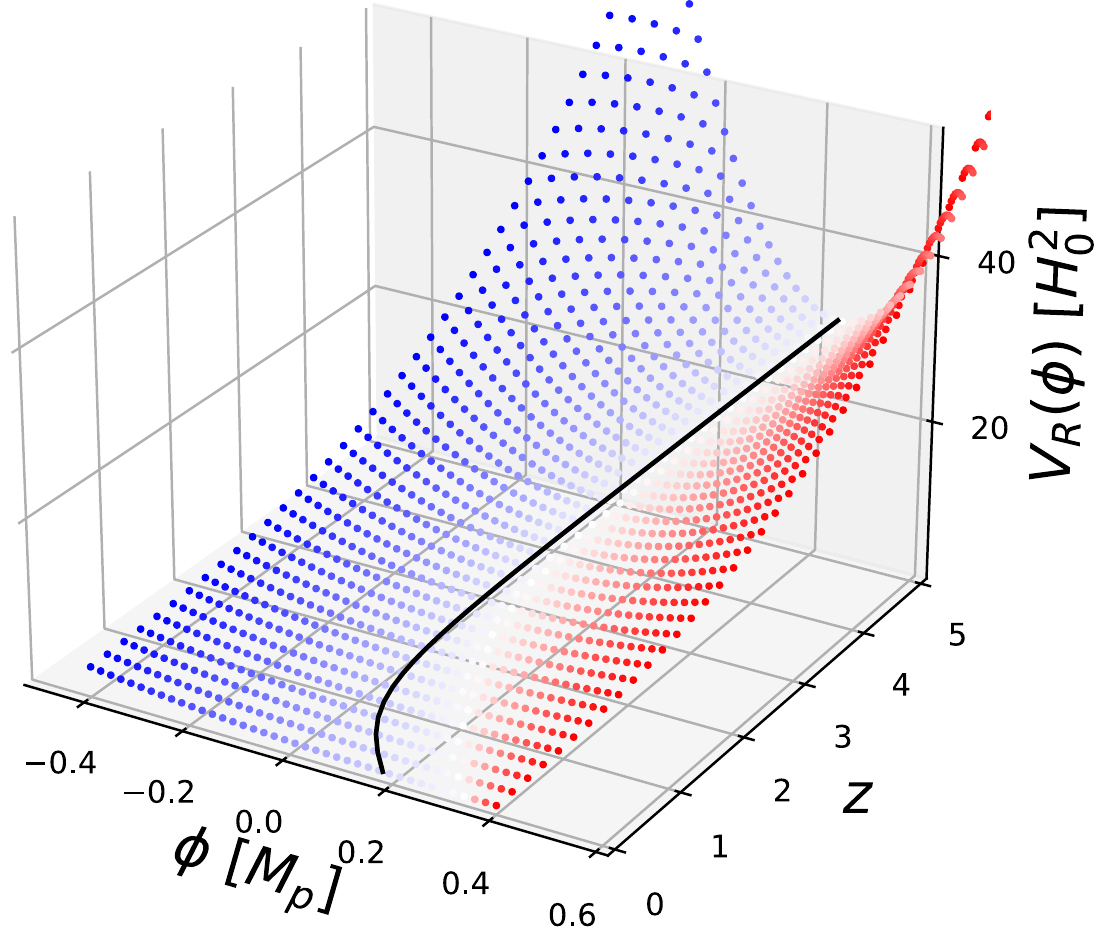}
    \includegraphics[width=0.53\linewidth]{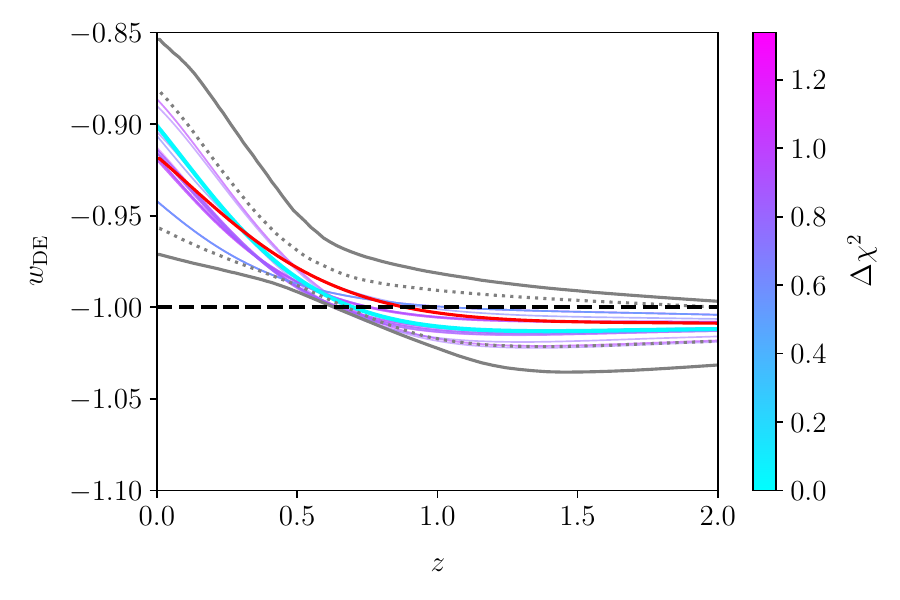}
    \caption{\textit{Left panel:} Shape of the curvature dependent effective potential $V_R(\phi)$ of the nonminimal coupling theory \eqref{eq:g4-L}, with $\xi=0.65, \lambda=1.4, V_0/3H_0^2=0.72$ taken from the bestfit cosmology. Potential surface is colored with respect to the strength of its gradient in the $\phi$ direction, with red for positive, blue for negative and white for $d V_R/d\phi=0$. Black solid line depicts the field trajectory output by the new \texttt{EFTCAMB}. \textit{Right panel:} Constraints on the functional shape of $w_{\rm{DE}}$ in the \textit{thawing gravity} model~\eqref{eq:g4-L}. The gray dotted and solid lines represent the 68\% and 95\% posterior regions, respectively. Black dashed line marks the position of $w_{\rm{DE}}=-1$. To further illustrate the $w_{\rm{DE}}$ shape in the DESI+CMB+SNIa constrained \textit{thawing gravity} model, we also plot 50 $w_{DE}$'s from the MCMC sample near the chain bestfit point in terms of $\chi^2$, with a color coding representing their $\chi^2$ difference $\Delta\chi^2=\chi^2-\chi^2_{\rm{bestfit}}$.}
    \label{fig:g4-model}
\end{figure*}
Based on the insight from previous discussion, we propose the following nonminimally coupled gravity theory
\begin{equation}\label{eq:g4-L}
	L=\frac{M_p^2}{2}\left[1-\xi (\phi/M_p)^2\right]R+X-V_0e^{-\lambda\phi/M_p},
\end{equation}
where $\{\xi,V_0,\lambda\}$ are constant parameters and we assume $\xi>0$ in this study. All species are nonminimally coupled to the same conformal factor in   Eq.\eqref{eq:g4-L}, but we stress that this is not necessarily the case in general, e.g. nonminimal coupling might be only in the dark sector. It can be checked that this model has a luminal tensor speed and is free of ghost and gradient instability.

Theory \eqref{eq:g4-L} provides rich dynamics with different combination of its three parameters $\{ V_0, \lambda, \xi \}$. Here we will only focus on the scenario that is of interest to explaining the DESI observation, which we dub as \textit{thawing gravity} for reasons becoming clear soon. Theory \eqref{eq:g4-L} has an effective potential $V_R(\phi) = \frac{1}{2}\xi R\phi^2+V_0e^{-\lambda\phi/M_p}$ with one global minimum for $\xi>0$. Initially when $R$ dominates,  
more specifically when $\xi M_p^2R\gg \lambda V_0$ and $\xi M_p^2R\gg \lambda^2 V_0$, the global minimum is close to zero
\begin{equation}\label{eq:g4-phisol}
	\phi_m\simeq\frac{\lambda V_0}{\xi M_p^2 R+\lambda^2V_0}M_p\sim0
\end{equation}
with ${m^2_{\phi}}\big|_{\phi=0} = \frac{d^2 V_R}{d\phi^2}= \xi M_p^2 R+\lambda^2V_0$. At early times (still in matter era), as long as $\xi$ is not too small, the $R$ dependent part of $m^2_{\phi}$ is much larger than $\lambda^2V_0$  (which approximates the DE energy scale), the field evolution is dominated by the quadratic part of $V_R$ and $\phi$ is localized at the global minimum $\phi_m$.
As the Universe expands and $R$ decreases, the second term in $m_{\phi}^2$ starts dominating and the field starts rolling along the scalar potential $V_0e^{-\lambda\phi/M_p}$, effectively thawing and deviating from $\phi_m$. This leads to an effective Planck mass $M^2_{\rm{eff}}\equiv(1-\xi\phi^2)M_p^2$ which decreases with time. The left panel of Fig.~\ref{fig:g4-model} further illustrates this dynamical picture by plotting $V_R[\phi(z)]$ as a two dimensional surface and the field evolution trajectory, for the bestfit model from our joint DESI+CMB+SNIa analysis.

Fitting the \textit{thawing gravity} \eqref{eq:g4-L} to DESI+CMB+SNIa, we plot the resulting $w_{\rm DE}\equiv\frac{ -2\dot{H}-3H^2-P_{\rm{m}} }{ 3H^2-\rho_{\rm{m}} }$, where the subscript ``m" refers to all species except for DE, in the right panel of Fig.~\ref{fig:g4-model}. We collect the detailed numerical setup and results in the Supplemental Material, but it is worth mentioning here that \textit{thawing gravity} only has two free parameters $\xi$ and $\lambda$, while $V_0$ is determined by the DE energy scale, which makes \textit{thawing gravity} having the same number of parameters as $w_0w_a$CDM. All of the plotted $w_{\rm{DE}}$ trends in Fig.~\ref{fig:g4-model}, as well as the mean functions, display phantom crossing behavior in $0.5<z<1$, consistent with what we observe in the nonparametric reconstruction. Fig.~\ref{fig:g4-model} indicates that for the \textit{thawing gravity} model the linear approximation of CPL works to some extent in the redshift range $0<z<1$. We thus derive the effective $w_0, w_a$ for each sampled point in the chain by least square fitting the actual $w_{\rm{DE}}(a)$ to the CPL parameterization in $0<z<1$ . The resulting $(w_0,w_a)$ posterior is shown in Fig.~\ref{fig:w0wa}; one can notice that it appears skewed, as expected, since it is associated to a linear approximation of the actual, theoretically derived $w_{\rm{DE}}$. As shown in Figs.~\ref{fig:w0wa} and \ref{fig:g4-model},  \textit{thawing gravity} provides a theoretical, covariant realization of the reconstructed EFT $\Omega$ as well as a natural way to implement phantom crossing in favor of DESI BAO, confirming conclusion obtained in the previous sections. \textit{thawing gravity} also fits better all data in the joint the dataset, with $\Delta\chi^2_{\rm{DESI}}=-2.1$, $\Delta\chi^2_{\rm{CMB}}=-1.8$ and $\Delta\chi^2_{\rm{SNIa}}=-1.9$ compared with $\Lambda$CDM, see the Supplemental Material for detail. Before concluding, we have to stress that these $\chi^2$ improvements are not enough to claim evidence for \textit{thawing gravity} over $\Lambda$CDM due to the additional parameters added.

{\bf Final Remarks - }%\section{Final remarks}
% The recent DESI DR1 BAO results, if confirmed, point to dynamical DE rather than a cosmological constant, with a $w_0w_a$CDM cosmology preferred at $\sim3\sigma$ over $\Lambda$CDM.
The recent BAO measurements form DESI DR1 reveal an inconsistency between CMB and BAO when analyzed within $\Lambda$CDM, leading to a $\sim3\sigma$ preference for dynamical dark energy in $w_0w_a$CDM.
In this \textit{letter} we showed the profound theoretical implications on DE and gravity of this finding. Firstly, we confirmed phantom crossing at $z<1$ as another important indication of the DESI observation, which rules out the most widely studied dark energy theory, quintessence. More importantly, our finding established that the DESI result, if confirmed, provides in fact the first observational hint of modified gravity, indicating that gravity is nonminimally coupled to matter contrary to what is postulated in GR. Based on these insights, we proposed a nonminimally coupled alternative theory of gravity, \textit{thawing gravity}, which naturally explains the phantom crossing while also fitting DESI BAO, CMB and SNIa better than $\Lambda$CDM.

\begin{acknowledgments}
GY thanks Yun-Song Piao for insightful discussions. The latest developer version (in preparation for release later this year) of \texttt{EFTCAMB}~\cite{Hu:2013twa,Raveri:2014cka}, based on \texttt{CAMB}~\cite{Lewis:1999bs}, is used to calculate the background and linear perturbation cosmology of all models studied in this \textit{letter}. The MCMC analysis is performed using the public cosmological sampler \texttt{Cobaya}~\cite{Torrado:2020dgo,2019ascl.soft10019T}. Some of the plots were made with \texttt{GetDist} \cite{Lewis:2019xzd}. The authors acknowledge computational support from the ALICE and Xmaris clusters of Leiden University. GY and AS acknowledge  support from the NWO and the Dutch Ministry of Education, Culture and Science (OCW) (through NWO VIDI Grant No. 2019/ENW/00678104 and ENW-XL Grant OCENW.XL21.XL21.025 DUSC) and from the D-ITP consortium. MM acknowledges funding by the Agenzia Spaziale Italiana (\textsc{asi}) under agreement no. 2018-23-HH.0 and support from INFN/Euclid Sezione di Roma. BH is supported by the National Natural Science Foundation of China Grants No. 12333001.
\end{acknowledgments}

\bibliography{reference}

%merlin.mbs apsrev4-1.bst 2010-07-25 4.21a (PWD, AO, DPC) hacked
%Control: key (0)
%Control: author (8) initials jnrlst
%Control: editor formatted (1) identically to author
%Control: production of article title (-1) disabled
%Control: page (0) single
%Control: year (1) truncated
%Control: production of eprint (0) enabled
\begin{thebibliography}{87}%
\makeatletter
\providecommand \@ifxundefined [1]{%
 \@ifx{#1\undefined}
}%
\providecommand \@ifnum [1]{%
 \ifnum #1\expandafter \@firstoftwo
 \else \expandafter \@secondoftwo
 \fi
}%
\providecommand \@ifx [1]{%
 \ifx #1\expandafter \@firstoftwo
 \else \expandafter \@secondoftwo
 \fi
}%
\providecommand \natexlab [1]{#1}%
\providecommand \enquote  [1]{``#1''}%
\providecommand \bibnamefont  [1]{#1}%
\providecommand \bibfnamefont [1]{#1}%
\providecommand \citenamefont [1]{#1}%
\providecommand \href@noop [0]{\@secondoftwo}%
\providecommand \href [0]{\begingroup \@sanitize@url \@href}%
\providecommand \@href[1]{\@@startlink{#1}\@@href}%
\providecommand \@@href[1]{\endgroup#1\@@endlink}%
\providecommand \@sanitize@url [0]{\catcode `\\12\catcode `\$12\catcode
  `\&12\catcode `\#12\catcode `\^12\catcode `\_12\catcode `\%12\relax}%
\providecommand \@@startlink[1]{}%
\providecommand \@@endlink[0]{}%
\providecommand \url  [0]{\begingroup\@sanitize@url \@url }%
\providecommand \@url [1]{\endgroup\@href {#1}{\urlprefix }}%
\providecommand \urlprefix  [0]{URL }%
\providecommand \Eprint [0]{\href }%
\providecommand \doibase [0]{http://dx.doi.org/}%
\providecommand \selectlanguage [0]{\@gobble}%
\providecommand \bibinfo  [0]{\@secondoftwo}%
\providecommand \bibfield  [0]{\@secondoftwo}%
\providecommand \translation [1]{[#1]}%
\providecommand \BibitemOpen [0]{}%
\providecommand \bibitemStop [0]{}%
\providecommand \bibitemNoStop [0]{.\EOS\space}%
\providecommand \EOS [0]{\spacefactor3000\relax}%
\providecommand \BibitemShut  [1]{\csname bibitem#1\endcsname}%
\let\auto@bib@innerbib\@empty
%</preamble>
\bibitem [{\citenamefont {Adame}\ \emph
  {et~al.}(2024{\natexlab{a}})\citenamefont {Adame} \emph
  {et~al.}}]{DESI:2024mwx}%
  \BibitemOpen
  \bibfield  {author} {\bibinfo {author} {\bibfnamefont {A.~G.}\ \bibnamefont
  {Adame}} \emph {et~al.} (\bibinfo {collaboration} {DESI}),\ }\href@noop {} {\
   (\bibinfo {year} {2024}{\natexlab{a}})},\ \Eprint
  {http://arxiv.org/abs/2404.03002} {arXiv:2404.03002 [astro-ph.CO]}
  \BibitemShut {NoStop}%
\bibitem [{\citenamefont {Adame}\ \emph
  {et~al.}(2024{\natexlab{b}})\citenamefont {Adame} \emph
  {et~al.}}]{DESI:2024uvr}%
  \BibitemOpen
  \bibfield  {author} {\bibinfo {author} {\bibfnamefont {A.~G.}\ \bibnamefont
  {Adame}} \emph {et~al.} (\bibinfo {collaboration} {DESI}),\ }\href@noop {} {\
   (\bibinfo {year} {2024}{\natexlab{b}})},\ \Eprint
  {http://arxiv.org/abs/2404.03000} {arXiv:2404.03000 [astro-ph.CO]}
  \BibitemShut {NoStop}%
\bibitem [{\citenamefont {Chevallier}\ and\ \citenamefont
  {Polarski}(2001)}]{Chevallier:2000qy}%
  \BibitemOpen
  \bibfield  {author} {\bibinfo {author} {\bibfnamefont {M.}~\bibnamefont
  {Chevallier}}\ and\ \bibinfo {author} {\bibfnamefont {D.}~\bibnamefont
  {Polarski}},\ }\href {\doibase 10.1142/S0218271801000822} {\bibfield
  {journal} {\bibinfo  {journal} {Int. J. Mod. Phys. D}\ }\textbf {\bibinfo
  {volume} {10}},\ \bibinfo {pages} {213} (\bibinfo {year} {2001})},\ \Eprint
  {http://arxiv.org/abs/gr-qc/0009008} {arXiv:gr-qc/0009008} \BibitemShut
  {NoStop}%
\bibitem [{\citenamefont {Linder}(2003)}]{Linder:2002et}%
  \BibitemOpen
  \bibfield  {author} {\bibinfo {author} {\bibfnamefont {E.~V.}\ \bibnamefont
  {Linder}},\ }\href {\doibase 10.1103/PhysRevLett.90.091301} {\bibfield
  {journal} {\bibinfo  {journal} {Phys. Rev. Lett.}\ }\textbf {\bibinfo
  {volume} {90}},\ \bibinfo {pages} {091301} (\bibinfo {year} {2003})},\
  \Eprint {http://arxiv.org/abs/astro-ph/0208512} {arXiv:astro-ph/0208512}
  \BibitemShut {NoStop}%
\bibitem [{\citenamefont {Scolnic}\ \emph {et~al.}(2022)\citenamefont {Scolnic}
  \emph {et~al.}}]{Scolnic:2021amr}%
  \BibitemOpen
  \bibfield  {author} {\bibinfo {author} {\bibfnamefont {D.}~\bibnamefont
  {Scolnic}} \emph {et~al.},\ }\href {\doibase 10.3847/1538-4357/ac8b7a}
  {\bibfield  {journal} {\bibinfo  {journal} {Astrophys. J.}\ }\textbf
  {\bibinfo {volume} {938}},\ \bibinfo {pages} {113} (\bibinfo {year}
  {2022})},\ \Eprint {http://arxiv.org/abs/2112.03863} {arXiv:2112.03863
  [astro-ph.CO]} \BibitemShut {NoStop}%
\bibitem [{\citenamefont {Rubin}\ \emph {et~al.}(2023)\citenamefont {Rubin}
  \emph {et~al.}}]{Rubin:2023ovl}%
  \BibitemOpen
  \bibfield  {author} {\bibinfo {author} {\bibfnamefont {D.}~\bibnamefont
  {Rubin}} \emph {et~al.},\ }\href@noop {} {\  (\bibinfo {year} {2023})},\
  \Eprint {http://arxiv.org/abs/2311.12098} {arXiv:2311.12098 [astro-ph.CO]}
  \BibitemShut {NoStop}%
\bibitem [{\citenamefont {Abbott}\ \emph {et~al.}(2024)\citenamefont {Abbott}
  \emph {et~al.}}]{DES:2024tys}%
  \BibitemOpen
  \bibfield  {author} {\bibinfo {author} {\bibfnamefont {T.~M.~C.}\
  \bibnamefont {Abbott}} \emph {et~al.} (\bibinfo {collaboration} {DES}),\
  }\href@noop {} {\  (\bibinfo {year} {2024})},\ \Eprint
  {http://arxiv.org/abs/2401.02929} {arXiv:2401.02929 [astro-ph.CO]}
  \BibitemShut {NoStop}%
\bibitem [{\citenamefont {Orchard}\ and\ \citenamefont
  {C\'ardenas}(2024)}]{Orchard:2024bve}%
  \BibitemOpen
  \bibfield  {author} {\bibinfo {author} {\bibfnamefont {L.}~\bibnamefont
  {Orchard}}\ and\ \bibinfo {author} {\bibfnamefont {V.~H.}\ \bibnamefont
  {C\'ardenas}},\ }\href@noop {} {\  (\bibinfo {year} {2024})},\ \Eprint
  {http://arxiv.org/abs/2407.05579} {arXiv:2407.05579 [astro-ph.CO]}
  \BibitemShut {NoStop}%
\bibitem [{\citenamefont {Chudaykin}\ and\ \citenamefont
  {Kunz}(2024)}]{Chudaykin:2024gol}%
  \BibitemOpen
  \bibfield  {author} {\bibinfo {author} {\bibfnamefont {A.}~\bibnamefont
  {Chudaykin}}\ and\ \bibinfo {author} {\bibfnamefont {M.}~\bibnamefont
  {Kunz}},\ }\href@noop {} {\  (\bibinfo {year} {2024})},\ \Eprint
  {http://arxiv.org/abs/2407.02558} {arXiv:2407.02558 [astro-ph.CO]}
  \BibitemShut {NoStop}%
\bibitem [{\citenamefont {Alestas}\ \emph {et~al.}(2024)\citenamefont
  {Alestas}, \citenamefont {Delgado}, \citenamefont {Ruiz}, \citenamefont
  {Akrami}, \citenamefont {Montero},\ and\ \citenamefont
  {Nesseris}}]{Alestas:2024gxe}%
  \BibitemOpen
  \bibfield  {author} {\bibinfo {author} {\bibfnamefont {G.}~\bibnamefont
  {Alestas}}, \bibinfo {author} {\bibfnamefont {M.}~\bibnamefont {Delgado}},
  \bibinfo {author} {\bibfnamefont {I.}~\bibnamefont {Ruiz}}, \bibinfo {author}
  {\bibfnamefont {Y.}~\bibnamefont {Akrami}}, \bibinfo {author} {\bibfnamefont
  {M.}~\bibnamefont {Montero}}, \ and\ \bibinfo {author} {\bibfnamefont
  {S.}~\bibnamefont {Nesseris}},\ }\href@noop {} {\  (\bibinfo {year}
  {2024})},\ \Eprint {http://arxiv.org/abs/2406.09212} {arXiv:2406.09212
  [hep-th]} \BibitemShut {NoStop}%
\bibitem [{\citenamefont {Wang}\ and\ \citenamefont
  {Piao}(2024)}]{Wang:2024dka}%
  \BibitemOpen
  \bibfield  {author} {\bibinfo {author} {\bibfnamefont {H.}~\bibnamefont
  {Wang}}\ and\ \bibinfo {author} {\bibfnamefont {Y.-S.}\ \bibnamefont
  {Piao}},\ }\href@noop {} {\  (\bibinfo {year} {2024})},\ \Eprint
  {http://arxiv.org/abs/2404.18579} {arXiv:2404.18579 [astro-ph.CO]}
  \BibitemShut {NoStop}%
\bibitem [{\citenamefont {Notari}\ \emph {et~al.}(2024)\citenamefont {Notari},
  \citenamefont {Redi},\ and\ \citenamefont {Tesi}}]{Notari:2024rti}%
  \BibitemOpen
  \bibfield  {author} {\bibinfo {author} {\bibfnamefont {A.}~\bibnamefont
  {Notari}}, \bibinfo {author} {\bibfnamefont {M.}~\bibnamefont {Redi}}, \ and\
  \bibinfo {author} {\bibfnamefont {A.}~\bibnamefont {Tesi}},\ }\href@noop {}
  {\  (\bibinfo {year} {2024})},\ \Eprint {http://arxiv.org/abs/2406.08459}
  {arXiv:2406.08459 [astro-ph.CO]} \BibitemShut {NoStop}%
\bibitem [{\citenamefont {Gialamas}\ \emph {et~al.}(2024)\citenamefont
  {Gialamas}, \citenamefont {H\"utsi}, \citenamefont {Kannike}, \citenamefont
  {Racioppi}, \citenamefont {Raidal}, \citenamefont {Vasar},\ and\
  \citenamefont {Veerm\"ae}}]{Gialamas:2024lyw}%
  \BibitemOpen
  \bibfield  {author} {\bibinfo {author} {\bibfnamefont {I.~D.}\ \bibnamefont
  {Gialamas}}, \bibinfo {author} {\bibfnamefont {G.}~\bibnamefont {H\"utsi}},
  \bibinfo {author} {\bibfnamefont {K.}~\bibnamefont {Kannike}}, \bibinfo
  {author} {\bibfnamefont {A.}~\bibnamefont {Racioppi}}, \bibinfo {author}
  {\bibfnamefont {M.}~\bibnamefont {Raidal}}, \bibinfo {author} {\bibfnamefont
  {M.}~\bibnamefont {Vasar}}, \ and\ \bibinfo {author} {\bibfnamefont
  {H.}~\bibnamefont {Veerm\"ae}},\ }\href@noop {} {\  (\bibinfo {year}
  {2024})},\ \Eprint {http://arxiv.org/abs/2406.07533} {arXiv:2406.07533
  [astro-ph.CO]} \BibitemShut {NoStop}%
\bibitem [{\citenamefont {Akarsu}\ \emph {et~al.}(2024)\citenamefont {Akarsu},
  \citenamefont {De~Felice}, \citenamefont {Di~Valentino}, \citenamefont
  {Kumar}, \citenamefont {Nunes}, \citenamefont {Ozulker}, \citenamefont
  {Vazquez},\ and\ \citenamefont {Yadav}}]{Akarsu:2024eoo}%
  \BibitemOpen
  \bibfield  {author} {\bibinfo {author} {\bibfnamefont {O.}~\bibnamefont
  {Akarsu}}, \bibinfo {author} {\bibfnamefont {A.}~\bibnamefont {De~Felice}},
  \bibinfo {author} {\bibfnamefont {E.}~\bibnamefont {Di~Valentino}}, \bibinfo
  {author} {\bibfnamefont {S.}~\bibnamefont {Kumar}}, \bibinfo {author}
  {\bibfnamefont {R.~C.}\ \bibnamefont {Nunes}}, \bibinfo {author}
  {\bibfnamefont {E.}~\bibnamefont {Ozulker}}, \bibinfo {author} {\bibfnamefont
  {J.~A.}\ \bibnamefont {Vazquez}}, \ and\ \bibinfo {author} {\bibfnamefont
  {A.}~\bibnamefont {Yadav}},\ }\href@noop {} {\  (\bibinfo {year} {2024})},\
  \Eprint {http://arxiv.org/abs/2406.07526} {arXiv:2406.07526 [astro-ph.CO]}
  \BibitemShut {NoStop}%
\bibitem [{\citenamefont {Heckman}\ \emph {et~al.}(2024)\citenamefont
  {Heckman}, \citenamefont {Ramadan},\ and\ \citenamefont
  {Sakstein}}]{Heckman:2024apk}%
  \BibitemOpen
  \bibfield  {author} {\bibinfo {author} {\bibfnamefont {J.~J.}\ \bibnamefont
  {Heckman}}, \bibinfo {author} {\bibfnamefont {O.~F.}\ \bibnamefont
  {Ramadan}}, \ and\ \bibinfo {author} {\bibfnamefont {J.}~\bibnamefont
  {Sakstein}},\ }\href@noop {} {\  (\bibinfo {year} {2024})},\ \Eprint
  {http://arxiv.org/abs/2406.04408} {arXiv:2406.04408 [astro-ph.CO]}
  \BibitemShut {NoStop}%
\bibitem [{\citenamefont {Ramadan}\ \emph {et~al.}(2024)\citenamefont
  {Ramadan}, \citenamefont {Sakstein},\ and\ \citenamefont
  {Rubin}}]{Ramadan:2024kmn}%
  \BibitemOpen
  \bibfield  {author} {\bibinfo {author} {\bibfnamefont {O.~F.}\ \bibnamefont
  {Ramadan}}, \bibinfo {author} {\bibfnamefont {J.}~\bibnamefont {Sakstein}}, \
  and\ \bibinfo {author} {\bibfnamefont {D.}~\bibnamefont {Rubin}},\
  }\href@noop {} {\  (\bibinfo {year} {2024})},\ \Eprint
  {http://arxiv.org/abs/2405.18747} {arXiv:2405.18747 [astro-ph.CO]}
  \BibitemShut {NoStop}%
\bibitem [{\citenamefont {Mukherjee}\ and\ \citenamefont
  {Sen}(2024)}]{Mukherjee:2024ryz}%
  \BibitemOpen
  \bibfield  {author} {\bibinfo {author} {\bibfnamefont {P.}~\bibnamefont
  {Mukherjee}}\ and\ \bibinfo {author} {\bibfnamefont {A.~A.}\ \bibnamefont
  {Sen}},\ }\href@noop {} {\  (\bibinfo {year} {2024})},\ \Eprint
  {http://arxiv.org/abs/2405.19178} {arXiv:2405.19178 [astro-ph.CO]}
  \BibitemShut {NoStop}%
\bibitem [{\citenamefont {Giar\`e}\ \emph {et~al.}(2024)\citenamefont
  {Giar\`e}, \citenamefont {Sabogal}, \citenamefont {Nunes},\ and\
  \citenamefont {Di~Valentino}}]{Giare:2024smz}%
  \BibitemOpen
  \bibfield  {author} {\bibinfo {author} {\bibfnamefont {W.}~\bibnamefont
  {Giar\`e}}, \bibinfo {author} {\bibfnamefont {M.~A.}\ \bibnamefont
  {Sabogal}}, \bibinfo {author} {\bibfnamefont {R.~C.}\ \bibnamefont {Nunes}},
  \ and\ \bibinfo {author} {\bibfnamefont {E.}~\bibnamefont {Di~Valentino}},\
  }\href@noop {} {\  (\bibinfo {year} {2024})},\ \Eprint
  {http://arxiv.org/abs/2404.15232} {arXiv:2404.15232 [astro-ph.CO]}
  \BibitemShut {NoStop}%
\bibitem [{\citenamefont {Berghaus}\ \emph {et~al.}(2024)\citenamefont
  {Berghaus}, \citenamefont {Kable},\ and\ \citenamefont
  {Miranda}}]{Berghaus:2024kra}%
  \BibitemOpen
  \bibfield  {author} {\bibinfo {author} {\bibfnamefont {K.~V.}\ \bibnamefont
  {Berghaus}}, \bibinfo {author} {\bibfnamefont {J.~A.}\ \bibnamefont {Kable}},
  \ and\ \bibinfo {author} {\bibfnamefont {V.}~\bibnamefont {Miranda}},\
  }\href@noop {} {\  (\bibinfo {year} {2024})},\ \Eprint
  {http://arxiv.org/abs/2404.14341} {arXiv:2404.14341 [astro-ph.CO]}
  \BibitemShut {NoStop}%
\bibitem [{\citenamefont {Yin}(2024)}]{Yin:2024hba}%
  \BibitemOpen
  \bibfield  {author} {\bibinfo {author} {\bibfnamefont {W.}~\bibnamefont
  {Yin}},\ }\href {\doibase 10.1007/JHEP05(2024)327} {\bibfield  {journal}
  {\bibinfo  {journal} {JHEP}\ }\textbf {\bibinfo {volume} {05}},\ \bibinfo
  {pages} {327} (\bibinfo {year} {2024})},\ \Eprint
  {http://arxiv.org/abs/2404.06444} {arXiv:2404.06444 [hep-ph]} \BibitemShut
  {NoStop}%
\bibitem [{\citenamefont {Tada}\ and\ \citenamefont
  {Terada}(2024)}]{Tada:2024znt}%
  \BibitemOpen
  \bibfield  {author} {\bibinfo {author} {\bibfnamefont {Y.}~\bibnamefont
  {Tada}}\ and\ \bibinfo {author} {\bibfnamefont {T.}~\bibnamefont {Terada}},\
  }\href {\doibase 10.1103/PhysRevD.109.L121305} {\bibfield  {journal}
  {\bibinfo  {journal} {Phys. Rev. D}\ }\textbf {\bibinfo {volume} {109}},\
  \bibinfo {pages} {L121305} (\bibinfo {year} {2024})},\ \Eprint
  {http://arxiv.org/abs/2404.05722} {arXiv:2404.05722 [astro-ph.CO]}
  \BibitemShut {NoStop}%
\bibitem [{\citenamefont {Bhattacharya}\ \emph {et~al.}(2024)\citenamefont
  {Bhattacharya}, \citenamefont {Borghetto}, \citenamefont {Malhotra},
  \citenamefont {Parameswaran}, \citenamefont {Tasinato},\ and\ \citenamefont
  {Zavala}}]{Bhattacharya:2024hep}%
  \BibitemOpen
  \bibfield  {author} {\bibinfo {author} {\bibfnamefont {S.}~\bibnamefont
  {Bhattacharya}}, \bibinfo {author} {\bibfnamefont {G.}~\bibnamefont
  {Borghetto}}, \bibinfo {author} {\bibfnamefont {A.}~\bibnamefont {Malhotra}},
  \bibinfo {author} {\bibfnamefont {S.}~\bibnamefont {Parameswaran}}, \bibinfo
  {author} {\bibfnamefont {G.}~\bibnamefont {Tasinato}}, \ and\ \bibinfo
  {author} {\bibfnamefont {I.}~\bibnamefont {Zavala}},\ }\href@noop {} {\
  (\bibinfo {year} {2024})},\ \Eprint {http://arxiv.org/abs/2405.17396}
  {arXiv:2405.17396 [astro-ph.CO]} \BibitemShut {NoStop}%
\bibitem [{\citenamefont {Dinda}(2024)}]{Dinda:2024kjf}%
  \BibitemOpen
  \bibfield  {author} {\bibinfo {author} {\bibfnamefont {B.~R.}\ \bibnamefont
  {Dinda}},\ }\href@noop {} {\  (\bibinfo {year} {2024})},\ \Eprint
  {http://arxiv.org/abs/2405.06618} {arXiv:2405.06618 [astro-ph.CO]}
  \BibitemShut {NoStop}%
\bibitem [{\citenamefont {Cort\^es}\ and\ \citenamefont
  {Liddle}(2024)}]{Cortes:2024lgw}%
  \BibitemOpen
  \bibfield  {author} {\bibinfo {author} {\bibfnamefont {M.}~\bibnamefont
  {Cort\^es}}\ and\ \bibinfo {author} {\bibfnamefont {A.~R.}\ \bibnamefont
  {Liddle}},\ }\href@noop {} {\  (\bibinfo {year} {2024})},\ \Eprint
  {http://arxiv.org/abs/2404.08056} {arXiv:2404.08056 [astro-ph.CO]}
  \BibitemShut {NoStop}%
\bibitem [{\citenamefont {Patel}\ and\ \citenamefont
  {Amendola}(2024)}]{Patel:2024odo}%
  \BibitemOpen
  \bibfield  {author} {\bibinfo {author} {\bibfnamefont {V.}~\bibnamefont
  {Patel}}\ and\ \bibinfo {author} {\bibfnamefont {L.}~\bibnamefont
  {Amendola}},\ }\href@noop {} {\  (\bibinfo {year} {2024})},\ \Eprint
  {http://arxiv.org/abs/2407.06586} {arXiv:2407.06586 [astro-ph.CO]}
  \BibitemShut {NoStop}%
\bibitem [{\citenamefont {Liu}\ \emph {et~al.}(2024)\citenamefont {Liu},
  \citenamefont {Wang},\ and\ \citenamefont {Zhao}}]{Liu:2024gfy}%
  \BibitemOpen
  \bibfield  {author} {\bibinfo {author} {\bibfnamefont {G.}~\bibnamefont
  {Liu}}, \bibinfo {author} {\bibfnamefont {Y.}~\bibnamefont {Wang}}, \ and\
  \bibinfo {author} {\bibfnamefont {W.}~\bibnamefont {Zhao}},\ }\href@noop {}
  {\  (\bibinfo {year} {2024})},\ \Eprint {http://arxiv.org/abs/2407.04385}
  {arXiv:2407.04385 [astro-ph.CO]} \BibitemShut {NoStop}%
\bibitem [{\citenamefont {Efstathiou}(2024)}]{Efstathiou:2024dvn}%
  \BibitemOpen
  \bibfield  {author} {\bibinfo {author} {\bibfnamefont {G.}~\bibnamefont
  {Efstathiou}}\ }(\bibinfo {year} {2024})\ \Eprint
  {http://arxiv.org/abs/2406.12106} {arXiv:2406.12106 [astro-ph.CO]}
  \BibitemShut {NoStop}%
\bibitem [{\citenamefont {Wang}(2024)}]{Wang:2024rjd}%
  \BibitemOpen
  \bibfield  {author} {\bibinfo {author} {\bibfnamefont {D.}~\bibnamefont
  {Wang}},\ }\href@noop {} {\  (\bibinfo {year} {2024})},\ \Eprint
  {http://arxiv.org/abs/2404.13833} {arXiv:2404.13833 [astro-ph.CO]}
  \BibitemShut {NoStop}%
\bibitem [{\citenamefont {Carloni}\ \emph {et~al.}(2024)\citenamefont
  {Carloni}, \citenamefont {Luongo},\ and\ \citenamefont
  {Muccino}}]{Carloni:2024zpl}%
  \BibitemOpen
  \bibfield  {author} {\bibinfo {author} {\bibfnamefont {Y.}~\bibnamefont
  {Carloni}}, \bibinfo {author} {\bibfnamefont {O.}~\bibnamefont {Luongo}}, \
  and\ \bibinfo {author} {\bibfnamefont {M.}~\bibnamefont {Muccino}},\
  }\href@noop {} {\  (\bibinfo {year} {2024})},\ \Eprint
  {http://arxiv.org/abs/2404.12068} {arXiv:2404.12068 [astro-ph.CO]}
  \BibitemShut {NoStop}%
\bibitem [{\citenamefont {Colg\'ain}\ \emph {et~al.}(2024)\citenamefont
  {Colg\'ain}, \citenamefont {Dainotti}, \citenamefont {Capozziello},
  \citenamefont {Pourojaghi}, \citenamefont {Sheikh-Jabbari},\ and\
  \citenamefont {Stojkovic}}]{Colgain:2024xqj}%
  \BibitemOpen
  \bibfield  {author} {\bibinfo {author} {\bibfnamefont {E.~O.}\ \bibnamefont
  {Colg\'ain}}, \bibinfo {author} {\bibfnamefont {M.~G.}\ \bibnamefont
  {Dainotti}}, \bibinfo {author} {\bibfnamefont {S.}~\bibnamefont
  {Capozziello}}, \bibinfo {author} {\bibfnamefont {S.}~\bibnamefont
  {Pourojaghi}}, \bibinfo {author} {\bibfnamefont {M.~M.}\ \bibnamefont
  {Sheikh-Jabbari}}, \ and\ \bibinfo {author} {\bibfnamefont {D.}~\bibnamefont
  {Stojkovic}},\ }\href@noop {} {\  (\bibinfo {year} {2024})},\ \Eprint
  {http://arxiv.org/abs/2404.08633} {arXiv:2404.08633 [astro-ph.CO]}
  \BibitemShut {NoStop}%
\bibitem [{\citenamefont {Luongo}\ and\ \citenamefont
  {Muccino}(2024)}]{Luongo:2024fww}%
  \BibitemOpen
  \bibfield  {author} {\bibinfo {author} {\bibfnamefont {O.}~\bibnamefont
  {Luongo}}\ and\ \bibinfo {author} {\bibfnamefont {M.}~\bibnamefont
  {Muccino}},\ }\href@noop {} {\  (\bibinfo {year} {2024})},\ \Eprint
  {http://arxiv.org/abs/2404.07070} {arXiv:2404.07070 [astro-ph.CO]}
  \BibitemShut {NoStop}%
\bibitem [{\citenamefont {Huang}\ \emph {et~al.}(2024)\citenamefont {Huang}
  \emph {et~al.}}]{Huang:2024qno}%
  \BibitemOpen
  \bibfield  {author} {\bibinfo {author} {\bibfnamefont {Z.}~\bibnamefont
  {Huang}} \emph {et~al.},\ }\href@noop {} {\  (\bibinfo {year} {2024})},\
  \Eprint {http://arxiv.org/abs/2405.03983} {arXiv:2405.03983 [astro-ph.CO]}
  \BibitemShut {NoStop}%
\bibitem [{\citenamefont {Jia}\ \emph {et~al.}(2024)\citenamefont {Jia},
  \citenamefont {Hu},\ and\ \citenamefont {Wang}}]{Jia:2024wix}%
  \BibitemOpen
  \bibfield  {author} {\bibinfo {author} {\bibfnamefont {X.~D.}\ \bibnamefont
  {Jia}}, \bibinfo {author} {\bibfnamefont {J.~P.}\ \bibnamefont {Hu}}, \ and\
  \bibinfo {author} {\bibfnamefont {F.~Y.}\ \bibnamefont {Wang}},\ }\href@noop
  {} {\  (\bibinfo {year} {2024})},\ \Eprint {http://arxiv.org/abs/2406.02019}
  {arXiv:2406.02019 [astro-ph.CO]} \BibitemShut {NoStop}%
\bibitem [{\citenamefont {Wang}\ \emph {et~al.}(2024)\citenamefont {Wang},
  \citenamefont {Lin}, \citenamefont {Ding},\ and\ \citenamefont
  {Hu}}]{Wang:2024pui}%
  \BibitemOpen
  \bibfield  {author} {\bibinfo {author} {\bibfnamefont {Z.}~\bibnamefont
  {Wang}}, \bibinfo {author} {\bibfnamefont {S.}~\bibnamefont {Lin}}, \bibinfo
  {author} {\bibfnamefont {Z.}~\bibnamefont {Ding}}, \ and\ \bibinfo {author}
  {\bibfnamefont {B.}~\bibnamefont {Hu}},\ }\href@noop {} {\  (\bibinfo {year}
  {2024})},\ \Eprint {http://arxiv.org/abs/2405.02168} {arXiv:2405.02168
  [astro-ph.CO]} \BibitemShut {NoStop}%
\bibitem [{\citenamefont {Shlivko}\ and\ \citenamefont
  {Steinhardt}(2024)}]{Shlivko:2024llw}%
  \BibitemOpen
  \bibfield  {author} {\bibinfo {author} {\bibfnamefont {D.}~\bibnamefont
  {Shlivko}}\ and\ \bibinfo {author} {\bibfnamefont {P.~J.}\ \bibnamefont
  {Steinhardt}},\ }\href {\doibase 10.1016/j.physletb.2024.138826} {\bibfield
  {journal} {\bibinfo  {journal} {Phys. Lett. B}\ }\textbf {\bibinfo {volume}
  {855}},\ \bibinfo {pages} {138826} (\bibinfo {year} {2024})},\ \Eprint
  {http://arxiv.org/abs/2405.03933} {arXiv:2405.03933 [astro-ph.CO]}
  \BibitemShut {NoStop}%
\bibitem [{\citenamefont {Creminelli}\ \emph {et~al.}(2009)\citenamefont
  {Creminelli}, \citenamefont {D'Amico}, \citenamefont {Norena},\ and\
  \citenamefont {Vernizzi}}]{Creminelli:2008wc}%
  \BibitemOpen
  \bibfield  {author} {\bibinfo {author} {\bibfnamefont {P.}~\bibnamefont
  {Creminelli}}, \bibinfo {author} {\bibfnamefont {G.}~\bibnamefont {D'Amico}},
  \bibinfo {author} {\bibfnamefont {J.}~\bibnamefont {Norena}}, \ and\ \bibinfo
  {author} {\bibfnamefont {F.}~\bibnamefont {Vernizzi}},\ }\href {\doibase
  10.1088/1475-7516/2009/02/018} {\bibfield  {journal} {\bibinfo  {journal}
  {JCAP}\ }\textbf {\bibinfo {volume} {02}},\ \bibinfo {pages} {018} (\bibinfo
  {year} {2009})},\ \Eprint {http://arxiv.org/abs/0811.0827} {arXiv:0811.0827
  [astro-ph]} \BibitemShut {NoStop}%
\bibitem [{\citenamefont {Gubitosi}\ \emph {et~al.}(2013)\citenamefont
  {Gubitosi}, \citenamefont {Piazza},\ and\ \citenamefont
  {Vernizzi}}]{Gubitosi:2012hu}%
  \BibitemOpen
  \bibfield  {author} {\bibinfo {author} {\bibfnamefont {G.}~\bibnamefont
  {Gubitosi}}, \bibinfo {author} {\bibfnamefont {F.}~\bibnamefont {Piazza}}, \
  and\ \bibinfo {author} {\bibfnamefont {F.}~\bibnamefont {Vernizzi}},\ }\href
  {\doibase 10.1088/1475-7516/2013/02/032} {\bibfield  {journal} {\bibinfo
  {journal} {JCAP}\ }\textbf {\bibinfo {volume} {02}},\ \bibinfo {pages} {032}
  (\bibinfo {year} {2013})},\ \Eprint {http://arxiv.org/abs/1210.0201}
  {arXiv:1210.0201 [hep-th]} \BibitemShut {NoStop}%
\bibitem [{\citenamefont {Bloomfield}\ \emph {et~al.}(2013)\citenamefont
  {Bloomfield}, \citenamefont {Flanagan}, \citenamefont {Park},\ and\
  \citenamefont {Watson}}]{Bloomfield:2012ff}%
  \BibitemOpen
  \bibfield  {author} {\bibinfo {author} {\bibfnamefont {J.~K.}\ \bibnamefont
  {Bloomfield}}, \bibinfo {author} {\bibfnamefont {E.~E.}\ \bibnamefont
  {Flanagan}}, \bibinfo {author} {\bibfnamefont {M.}~\bibnamefont {Park}}, \
  and\ \bibinfo {author} {\bibfnamefont {S.}~\bibnamefont {Watson}},\ }\href
  {\doibase 10.1088/1475-7516/2013/08/010} {\bibfield  {journal} {\bibinfo
  {journal} {JCAP}\ }\textbf {\bibinfo {volume} {08}},\ \bibinfo {pages} {010}
  (\bibinfo {year} {2013})},\ \Eprint {http://arxiv.org/abs/1211.7054}
  {arXiv:1211.7054 [astro-ph.CO]} \BibitemShut {NoStop}%
\bibitem [{\citenamefont {Horndeski}(1974)}]{Horndeski:1974wa}%
  \BibitemOpen
  \bibfield  {author} {\bibinfo {author} {\bibfnamefont {G.~W.}\ \bibnamefont
  {Horndeski}},\ }\href {\doibase 10.1007/BF01807638} {\bibfield  {journal}
  {\bibinfo  {journal} {Int. J. Theor. Phys.}\ }\textbf {\bibinfo {volume}
  {10}},\ \bibinfo {pages} {363} (\bibinfo {year} {1974})}\BibitemShut
  {NoStop}%
%%CITATION = IJTPB,10,363;%%
\bibitem [{\citenamefont {Gleyzes}\ \emph {et~al.}(2013)\citenamefont
  {Gleyzes}, \citenamefont {Langlois}, \citenamefont {Piazza},\ and\
  \citenamefont {Vernizzi}}]{Gleyzes:2013ooa}%
  \BibitemOpen
  \bibfield  {author} {\bibinfo {author} {\bibfnamefont {J.}~\bibnamefont
  {Gleyzes}}, \bibinfo {author} {\bibfnamefont {D.}~\bibnamefont {Langlois}},
  \bibinfo {author} {\bibfnamefont {F.}~\bibnamefont {Piazza}}, \ and\ \bibinfo
  {author} {\bibfnamefont {F.}~\bibnamefont {Vernizzi}},\ }\href {\doibase
  10.1088/1475-7516/2013/08/025} {\bibfield  {journal} {\bibinfo  {journal}
  {JCAP}\ }\textbf {\bibinfo {volume} {08}},\ \bibinfo {pages} {025} (\bibinfo
  {year} {2013})},\ \Eprint {http://arxiv.org/abs/1304.4840} {arXiv:1304.4840
  [hep-th]} \BibitemShut {NoStop}%
\bibitem [{\citenamefont {Bloomfield}(2013)}]{Bloomfield:2013efa}%
  \BibitemOpen
  \bibfield  {author} {\bibinfo {author} {\bibfnamefont {J.}~\bibnamefont
  {Bloomfield}},\ }\href {\doibase 10.1088/1475-7516/2013/12/044} {\bibfield
  {journal} {\bibinfo  {journal} {JCAP}\ }\textbf {\bibinfo {volume} {12}},\
  \bibinfo {pages} {044} (\bibinfo {year} {2013})},\ \Eprint
  {http://arxiv.org/abs/1304.6712} {arXiv:1304.6712 [astro-ph.CO]} \BibitemShut
  {NoStop}%
\bibitem [{\citenamefont {Kennedy}\ \emph {et~al.}(2017)\citenamefont
  {Kennedy}, \citenamefont {Lombriser},\ and\ \citenamefont
  {Taylor}}]{Kennedy:2017sof}%
  \BibitemOpen
  \bibfield  {author} {\bibinfo {author} {\bibfnamefont {J.}~\bibnamefont
  {Kennedy}}, \bibinfo {author} {\bibfnamefont {L.}~\bibnamefont {Lombriser}},
  \ and\ \bibinfo {author} {\bibfnamefont {A.}~\bibnamefont {Taylor}},\ }\href
  {\doibase 10.1103/PhysRevD.96.084051} {\bibfield  {journal} {\bibinfo
  {journal} {Phys. Rev. D}\ }\textbf {\bibinfo {volume} {96}},\ \bibinfo
  {pages} {084051} (\bibinfo {year} {2017})},\ \Eprint
  {http://arxiv.org/abs/1705.09290} {arXiv:1705.09290 [gr-qc]} \BibitemShut
  {NoStop}%
\bibitem [{\citenamefont {Gelman}\ and\ \citenamefont
  {Rubin}(1992)}]{Gelman:1992zz}%
  \BibitemOpen
  \bibfield  {author} {\bibinfo {author} {\bibfnamefont {A.}~\bibnamefont
  {Gelman}}\ and\ \bibinfo {author} {\bibfnamefont {D.~B.}\ \bibnamefont
  {Rubin}},\ }\href {\doibase 10.1214/ss/1177011136} {\bibfield  {journal}
  {\bibinfo  {journal} {Statist. Sci.}\ }\textbf {\bibinfo {volume} {7}},\
  \bibinfo {pages} {457} (\bibinfo {year} {1992})}\BibitemShut {NoStop}%
\bibitem [{\citenamefont {Bennett}\ \emph {et~al.}(2021)\citenamefont
  {Bennett}, \citenamefont {Buldgen}, \citenamefont {De~Salas}, \citenamefont
  {Drewes}, \citenamefont {Gariazzo}, \citenamefont {Pastor},\ and\
  \citenamefont {Wong}}]{Bennett:2020zkv}%
  \BibitemOpen
  \bibfield  {author} {\bibinfo {author} {\bibfnamefont {J.~J.}\ \bibnamefont
  {Bennett}}, \bibinfo {author} {\bibfnamefont {G.}~\bibnamefont {Buldgen}},
  \bibinfo {author} {\bibfnamefont {P.~F.}\ \bibnamefont {De~Salas}}, \bibinfo
  {author} {\bibfnamefont {M.}~\bibnamefont {Drewes}}, \bibinfo {author}
  {\bibfnamefont {S.}~\bibnamefont {Gariazzo}}, \bibinfo {author}
  {\bibfnamefont {S.}~\bibnamefont {Pastor}}, \ and\ \bibinfo {author}
  {\bibfnamefont {Y.~Y.~Y.}\ \bibnamefont {Wong}},\ }\href {\doibase
  10.1088/1475-7516/2021/04/073} {\bibfield  {journal} {\bibinfo  {journal}
  {JCAP}\ }\textbf {\bibinfo {volume} {04}},\ \bibinfo {pages} {073} (\bibinfo
  {year} {2021})},\ \Eprint {http://arxiv.org/abs/2012.02726} {arXiv:2012.02726
  [hep-ph]} \BibitemShut {NoStop}%
\bibitem [{\citenamefont {Froustey}\ \emph {et~al.}(2020)\citenamefont
  {Froustey}, \citenamefont {Pitrou},\ and\ \citenamefont
  {Volpe}}]{Froustey:2020mcq}%
  \BibitemOpen
  \bibfield  {author} {\bibinfo {author} {\bibfnamefont {J.}~\bibnamefont
  {Froustey}}, \bibinfo {author} {\bibfnamefont {C.}~\bibnamefont {Pitrou}}, \
  and\ \bibinfo {author} {\bibfnamefont {M.~C.}\ \bibnamefont {Volpe}},\ }\href
  {\doibase 10.1088/1475-7516/2020/12/015} {\bibfield  {journal} {\bibinfo
  {journal} {JCAP}\ }\textbf {\bibinfo {volume} {12}},\ \bibinfo {pages} {015}
  (\bibinfo {year} {2020})},\ \Eprint {http://arxiv.org/abs/2008.01074}
  {arXiv:2008.01074 [hep-ph]} \BibitemShut {NoStop}%
\bibitem [{\citenamefont {Akita}\ and\ \citenamefont
  {Yamaguchi}(2020)}]{Akita:2020szl}%
  \BibitemOpen
  \bibfield  {author} {\bibinfo {author} {\bibfnamefont {K.}~\bibnamefont
  {Akita}}\ and\ \bibinfo {author} {\bibfnamefont {M.}~\bibnamefont
  {Yamaguchi}},\ }\href {\doibase 10.1088/1475-7516/2020/08/012} {\bibfield
  {journal} {\bibinfo  {journal} {JCAP}\ }\textbf {\bibinfo {volume} {08}},\
  \bibinfo {pages} {012} (\bibinfo {year} {2020})},\ \Eprint
  {http://arxiv.org/abs/2005.07047} {arXiv:2005.07047 [hep-ph]} \BibitemShut
  {NoStop}%
\bibitem [{\citenamefont {Frusciante}\ \emph {et~al.}(2016)\citenamefont
  {Frusciante}, \citenamefont {Papadomanolakis},\ and\ \citenamefont
  {Silvestri}}]{Frusciante:2016xoj}%
  \BibitemOpen
  \bibfield  {author} {\bibinfo {author} {\bibfnamefont {N.}~\bibnamefont
  {Frusciante}}, \bibinfo {author} {\bibfnamefont {G.}~\bibnamefont
  {Papadomanolakis}}, \ and\ \bibinfo {author} {\bibfnamefont {A.}~\bibnamefont
  {Silvestri}},\ }\href {\doibase 10.1088/1475-7516/2016/07/018} {\bibfield
  {journal} {\bibinfo  {journal} {JCAP}\ }\textbf {\bibinfo {volume} {07}},\
  \bibinfo {pages} {018} (\bibinfo {year} {2016})},\ \Eprint
  {http://arxiv.org/abs/1601.04064} {arXiv:1601.04064 [gr-qc]} \BibitemShut
  {NoStop}%
\bibitem [{\citenamefont {De~Felice}\ \emph {et~al.}(2017)\citenamefont
  {De~Felice}, \citenamefont {Frusciante},\ and\ \citenamefont
  {Papadomanolakis}}]{DeFelice:2016ucp}%
  \BibitemOpen
  \bibfield  {author} {\bibinfo {author} {\bibfnamefont {A.}~\bibnamefont
  {De~Felice}}, \bibinfo {author} {\bibfnamefont {N.}~\bibnamefont
  {Frusciante}}, \ and\ \bibinfo {author} {\bibfnamefont {G.}~\bibnamefont
  {Papadomanolakis}},\ }\href {\doibase 10.1088/1475-7516/2017/03/027}
  {\bibfield  {journal} {\bibinfo  {journal} {JCAP}\ }\textbf {\bibinfo
  {volume} {03}},\ \bibinfo {pages} {027} (\bibinfo {year} {2017})},\ \Eprint
  {http://arxiv.org/abs/1609.03599} {arXiv:1609.03599 [gr-qc]} \BibitemShut
  {NoStop}%
\bibitem [{\citenamefont {Peirone}\ \emph {et~al.}(2017)\citenamefont
  {Peirone}, \citenamefont {Martinelli}, \citenamefont {Raveri},\ and\
  \citenamefont {Silvestri}}]{Peirone:2017lgi}%
  \BibitemOpen
  \bibfield  {author} {\bibinfo {author} {\bibfnamefont {S.}~\bibnamefont
  {Peirone}}, \bibinfo {author} {\bibfnamefont {M.}~\bibnamefont {Martinelli}},
  \bibinfo {author} {\bibfnamefont {M.}~\bibnamefont {Raveri}}, \ and\ \bibinfo
  {author} {\bibfnamefont {A.}~\bibnamefont {Silvestri}},\ }\href {\doibase
  10.1103/PhysRevD.96.063524} {\bibfield  {journal} {\bibinfo  {journal} {Phys.
  Rev. D}\ }\textbf {\bibinfo {volume} {96}},\ \bibinfo {pages} {063524}
  (\bibinfo {year} {2017})},\ \Eprint {http://arxiv.org/abs/1702.06526}
  {arXiv:1702.06526 [astro-ph.CO]} \BibitemShut {NoStop}%
\bibitem [{\citenamefont {Gerardi}\ \emph {et~al.}(2019)\citenamefont
  {Gerardi}, \citenamefont {Martinelli},\ and\ \citenamefont
  {Silvestri}}]{Gerardi:2019obr}%
  \BibitemOpen
  \bibfield  {author} {\bibinfo {author} {\bibfnamefont {F.}~\bibnamefont
  {Gerardi}}, \bibinfo {author} {\bibfnamefont {M.}~\bibnamefont {Martinelli}},
  \ and\ \bibinfo {author} {\bibfnamefont {A.}~\bibnamefont {Silvestri}},\
  }\href {\doibase 10.1088/1475-7516/2019/07/042} {\bibfield  {journal}
  {\bibinfo  {journal} {JCAP}\ }\textbf {\bibinfo {volume} {07}},\ \bibinfo
  {pages} {042} (\bibinfo {year} {2019})},\ \Eprint
  {http://arxiv.org/abs/1902.09423} {arXiv:1902.09423 [astro-ph.CO]}
  \BibitemShut {NoStop}%
\bibitem [{\citenamefont {Frusciante}\ \emph
  {et~al.}(2019{\natexlab{a}})\citenamefont {Frusciante}, \citenamefont
  {Papadomanolakis}, \citenamefont {Peirone},\ and\ \citenamefont
  {Silvestri}}]{Frusciante:2018vht}%
  \BibitemOpen
  \bibfield  {author} {\bibinfo {author} {\bibfnamefont {N.}~\bibnamefont
  {Frusciante}}, \bibinfo {author} {\bibfnamefont {G.}~\bibnamefont
  {Papadomanolakis}}, \bibinfo {author} {\bibfnamefont {S.}~\bibnamefont
  {Peirone}}, \ and\ \bibinfo {author} {\bibfnamefont {A.}~\bibnamefont
  {Silvestri}},\ }\href {\doibase 10.1088/1475-7516/2019/02/029} {\bibfield
  {journal} {\bibinfo  {journal} {JCAP}\ }\textbf {\bibinfo {volume} {02}},\
  \bibinfo {pages} {029} (\bibinfo {year} {2019}{\natexlab{a}})},\ \Eprint
  {http://arxiv.org/abs/1810.03461} {arXiv:1810.03461 [gr-qc]} \BibitemShut
  {NoStop}%
\bibitem [{\citenamefont {Peirone}\ \emph {et~al.}(2018)\citenamefont
  {Peirone}, \citenamefont {Koyama}, \citenamefont {Pogosian}, \citenamefont
  {Raveri},\ and\ \citenamefont {Silvestri}}]{Peirone:2017ywi}%
  \BibitemOpen
  \bibfield  {author} {\bibinfo {author} {\bibfnamefont {S.}~\bibnamefont
  {Peirone}}, \bibinfo {author} {\bibfnamefont {K.}~\bibnamefont {Koyama}},
  \bibinfo {author} {\bibfnamefont {L.}~\bibnamefont {Pogosian}}, \bibinfo
  {author} {\bibfnamefont {M.}~\bibnamefont {Raveri}}, \ and\ \bibinfo {author}
  {\bibfnamefont {A.}~\bibnamefont {Silvestri}},\ }\href {\doibase
  10.1103/PhysRevD.97.043519} {\bibfield  {journal} {\bibinfo  {journal} {Phys.
  Rev. D}\ }\textbf {\bibinfo {volume} {97}},\ \bibinfo {pages} {043519}
  (\bibinfo {year} {2018})},\ \Eprint {http://arxiv.org/abs/1712.00444}
  {arXiv:1712.00444 [astro-ph.CO]} \BibitemShut {NoStop}%
\bibitem [{\citenamefont {Espejo}\ \emph {et~al.}(2019)\citenamefont {Espejo},
  \citenamefont {Peirone}, \citenamefont {Raveri}, \citenamefont {Koyama},
  \citenamefont {Pogosian},\ and\ \citenamefont {Silvestri}}]{Espejo:2018hxa}%
  \BibitemOpen
  \bibfield  {author} {\bibinfo {author} {\bibfnamefont {J.}~\bibnamefont
  {Espejo}}, \bibinfo {author} {\bibfnamefont {S.}~\bibnamefont {Peirone}},
  \bibinfo {author} {\bibfnamefont {M.}~\bibnamefont {Raveri}}, \bibinfo
  {author} {\bibfnamefont {K.}~\bibnamefont {Koyama}}, \bibinfo {author}
  {\bibfnamefont {L.}~\bibnamefont {Pogosian}}, \ and\ \bibinfo {author}
  {\bibfnamefont {A.}~\bibnamefont {Silvestri}},\ }\href {\doibase
  10.1103/PhysRevD.99.023512} {\bibfield  {journal} {\bibinfo  {journal} {Phys.
  Rev. D}\ }\textbf {\bibinfo {volume} {99}},\ \bibinfo {pages} {023512}
  (\bibinfo {year} {2019})},\ \Eprint {http://arxiv.org/abs/1809.01121}
  {arXiv:1809.01121 [astro-ph.CO]} \BibitemShut {NoStop}%
\bibitem [{\citenamefont {Adams}\ \emph {et~al.}(2006)\citenamefont {Adams},
  \citenamefont {Arkani-Hamed}, \citenamefont {Dubovsky}, \citenamefont
  {Nicolis},\ and\ \citenamefont {Rattazzi}}]{Adams:2006sv}%
  \BibitemOpen
  \bibfield  {author} {\bibinfo {author} {\bibfnamefont {A.}~\bibnamefont
  {Adams}}, \bibinfo {author} {\bibfnamefont {N.}~\bibnamefont {Arkani-Hamed}},
  \bibinfo {author} {\bibfnamefont {S.}~\bibnamefont {Dubovsky}}, \bibinfo
  {author} {\bibfnamefont {A.}~\bibnamefont {Nicolis}}, \ and\ \bibinfo
  {author} {\bibfnamefont {R.}~\bibnamefont {Rattazzi}},\ }\href {\doibase
  10.1088/1126-6708/2006/10/014} {\bibfield  {journal} {\bibinfo  {journal}
  {JHEP}\ }\textbf {\bibinfo {volume} {10}},\ \bibinfo {pages} {014} (\bibinfo
  {year} {2006})},\ \Eprint {http://arxiv.org/abs/hep-th/0602178}
  {arXiv:hep-th/0602178} \BibitemShut {NoStop}%
\bibitem [{\citenamefont {Raveri}(2020)}]{Raveri:2019mxg}%
  \BibitemOpen
  \bibfield  {author} {\bibinfo {author} {\bibfnamefont {M.}~\bibnamefont
  {Raveri}},\ }\href {\doibase 10.1103/PhysRevD.101.083524} {\bibfield
  {journal} {\bibinfo  {journal} {Phys. Rev. D}\ }\textbf {\bibinfo {volume}
  {101}},\ \bibinfo {pages} {083524} (\bibinfo {year} {2020})},\ \Eprint
  {http://arxiv.org/abs/1902.01366} {arXiv:1902.01366 [astro-ph.CO]}
  \BibitemShut {NoStop}%
\bibitem [{\citenamefont {Frusciante}\ \emph
  {et~al.}(2019{\natexlab{b}})\citenamefont {Frusciante}, \citenamefont
  {Peirone}, \citenamefont {Casas},\ and\ \citenamefont
  {Lima}}]{Frusciante:2018jzw}%
  \BibitemOpen
  \bibfield  {author} {\bibinfo {author} {\bibfnamefont {N.}~\bibnamefont
  {Frusciante}}, \bibinfo {author} {\bibfnamefont {S.}~\bibnamefont {Peirone}},
  \bibinfo {author} {\bibfnamefont {S.}~\bibnamefont {Casas}}, \ and\ \bibinfo
  {author} {\bibfnamefont {N.~A.}\ \bibnamefont {Lima}},\ }\href {\doibase
  10.1103/PhysRevD.99.063538} {\bibfield  {journal} {\bibinfo  {journal} {Phys.
  Rev. D}\ }\textbf {\bibinfo {volume} {99}},\ \bibinfo {pages} {063538}
  (\bibinfo {year} {2019}{\natexlab{b}})},\ \Eprint
  {http://arxiv.org/abs/1810.10521} {arXiv:1810.10521 [astro-ph.CO]}
  \BibitemShut {NoStop}%
\bibitem [{\citenamefont {Deffayet}\ \emph {et~al.}(2010)\citenamefont
  {Deffayet}, \citenamefont {Pujolas}, \citenamefont {Sawicki},\ and\
  \citenamefont {Vikman}}]{Deffayet:2010qz}%
  \BibitemOpen
  \bibfield  {author} {\bibinfo {author} {\bibfnamefont {C.}~\bibnamefont
  {Deffayet}}, \bibinfo {author} {\bibfnamefont {O.}~\bibnamefont {Pujolas}},
  \bibinfo {author} {\bibfnamefont {I.}~\bibnamefont {Sawicki}}, \ and\
  \bibinfo {author} {\bibfnamefont {A.}~\bibnamefont {Vikman}},\ }\href
  {\doibase 10.1088/1475-7516/2010/10/026} {\bibfield  {journal} {\bibinfo
  {journal} {JCAP}\ }\textbf {\bibinfo {volume} {10}},\ \bibinfo {pages} {026}
  (\bibinfo {year} {2010})},\ \Eprint {http://arxiv.org/abs/1008.0048}
  {arXiv:1008.0048 [hep-th]} \BibitemShut {NoStop}%
\bibitem [{\citenamefont {de~Boe}\ \emph {et~al.}(2024)\citenamefont {de~Boe},
  \citenamefont {Ye}, \citenamefont {Renzi}, \citenamefont {Albuquerque},
  \citenamefont {Frusciante},\ and\ \citenamefont {Silvestri}}]{deBoe:2024gpf}%
  \BibitemOpen
  \bibfield  {author} {\bibinfo {author} {\bibfnamefont {D.}~\bibnamefont
  {de~Boe}}, \bibinfo {author} {\bibfnamefont {G.}~\bibnamefont {Ye}}, \bibinfo
  {author} {\bibfnamefont {F.}~\bibnamefont {Renzi}}, \bibinfo {author}
  {\bibfnamefont {I.~S.}\ \bibnamefont {Albuquerque}}, \bibinfo {author}
  {\bibfnamefont {N.}~\bibnamefont {Frusciante}}, \ and\ \bibinfo {author}
  {\bibfnamefont {A.}~\bibnamefont {Silvestri}},\ }\href@noop {} {\  (\bibinfo
  {year} {2024})},\ \Eprint {http://arxiv.org/abs/2403.13096} {arXiv:2403.13096
  [astro-ph.CO]} \BibitemShut {NoStop}%
\bibitem [{\citenamefont {Ross}\ \emph {et~al.}(2015)\citenamefont {Ross},
  \citenamefont {Samushia}, \citenamefont {Howlett}, \citenamefont {Percival},
  \citenamefont {Burden},\ and\ \citenamefont {Manera}}]{Ross:2014qpa}%
  \BibitemOpen
  \bibfield  {author} {\bibinfo {author} {\bibfnamefont {A.~J.}\ \bibnamefont
  {Ross}}, \bibinfo {author} {\bibfnamefont {L.}~\bibnamefont {Samushia}},
  \bibinfo {author} {\bibfnamefont {C.}~\bibnamefont {Howlett}}, \bibinfo
  {author} {\bibfnamefont {W.~J.}\ \bibnamefont {Percival}}, \bibinfo {author}
  {\bibfnamefont {A.}~\bibnamefont {Burden}}, \ and\ \bibinfo {author}
  {\bibfnamefont {M.}~\bibnamefont {Manera}},\ }\href {\doibase
  10.1093/mnras/stv154} {\bibfield  {journal} {\bibinfo  {journal} {Mon. Not.
  Roy. Astron. Soc.}\ }\textbf {\bibinfo {volume} {449}},\ \bibinfo {pages}
  {835} (\bibinfo {year} {2015})},\ \Eprint {http://arxiv.org/abs/1409.3242}
  {arXiv:1409.3242 [astro-ph.CO]} \BibitemShut {NoStop}%
\bibitem [{\citenamefont {Alam}\ \emph {et~al.}(2017)\citenamefont {Alam} \emph
  {et~al.}}]{BOSS:2016wmc}%
  \BibitemOpen
  \bibfield  {author} {\bibinfo {author} {\bibfnamefont {S.}~\bibnamefont
  {Alam}} \emph {et~al.} (\bibinfo {collaboration} {BOSS}),\ }\href {\doibase
  10.1093/mnras/stx721} {\bibfield  {journal} {\bibinfo  {journal} {Mon. Not.
  Roy. Astron. Soc.}\ }\textbf {\bibinfo {volume} {470}},\ \bibinfo {pages}
  {2617} (\bibinfo {year} {2017})},\ \Eprint {http://arxiv.org/abs/1607.03155}
  {arXiv:1607.03155 [astro-ph.CO]} \BibitemShut {NoStop}%
\bibitem [{\citenamefont {Alam}\ \emph {et~al.}(2021)\citenamefont {Alam} \emph
  {et~al.}}]{eBOSS:2020yzd}%
  \BibitemOpen
  \bibfield  {author} {\bibinfo {author} {\bibfnamefont {S.}~\bibnamefont
  {Alam}} \emph {et~al.} (\bibinfo {collaboration} {eBOSS}),\ }\href {\doibase
  10.1103/PhysRevD.103.083533} {\bibfield  {journal} {\bibinfo  {journal}
  {Phys. Rev. D}\ }\textbf {\bibinfo {volume} {103}},\ \bibinfo {pages}
  {083533} (\bibinfo {year} {2021})},\ \Eprint
  {http://arxiv.org/abs/2007.08991} {arXiv:2007.08991 [astro-ph.CO]}
  \BibitemShut {NoStop}%
\bibitem [{\citenamefont {Asgari}\ \emph {et~al.}(2021)\citenamefont {Asgari}
  \emph {et~al.}}]{KiDS:2020suj}%
  \BibitemOpen
  \bibfield  {author} {\bibinfo {author} {\bibfnamefont {M.}~\bibnamefont
  {Asgari}} \emph {et~al.} (\bibinfo {collaboration} {KiDS}),\ }\href {\doibase
  10.1051/0004-6361/202039070} {\bibfield  {journal} {\bibinfo  {journal}
  {Astron. Astrophys.}\ }\textbf {\bibinfo {volume} {645}},\ \bibinfo {pages}
  {A104} (\bibinfo {year} {2021})},\ \Eprint {http://arxiv.org/abs/2007.15633}
  {arXiv:2007.15633 [astro-ph.CO]} \BibitemShut {NoStop}%
\bibitem [{\citenamefont {Amon}\ \emph {et~al.}(2022)\citenamefont {Amon} \emph
  {et~al.}}]{DES:2021bvc}%
  \BibitemOpen
  \bibfield  {author} {\bibinfo {author} {\bibfnamefont {A.}~\bibnamefont
  {Amon}} \emph {et~al.} (\bibinfo {collaboration} {DES}),\ }\href {\doibase
  10.1103/PhysRevD.105.023514} {\bibfield  {journal} {\bibinfo  {journal}
  {Phys. Rev. D}\ }\textbf {\bibinfo {volume} {105}},\ \bibinfo {pages}
  {023514} (\bibinfo {year} {2022})},\ \Eprint
  {http://arxiv.org/abs/2105.13543} {arXiv:2105.13543 [astro-ph.CO]}
  \BibitemShut {NoStop}%
\bibitem [{\citenamefont {Secco}\ \emph {et~al.}(2022)\citenamefont {Secco}
  \emph {et~al.}}]{DES:2021vln}%
  \BibitemOpen
  \bibfield  {author} {\bibinfo {author} {\bibfnamefont {L.~F.}\ \bibnamefont
  {Secco}} \emph {et~al.} (\bibinfo {collaboration} {DES}),\ }\href {\doibase
  10.1103/PhysRevD.105.023515} {\bibfield  {journal} {\bibinfo  {journal}
  {Phys. Rev. D}\ }\textbf {\bibinfo {volume} {105}},\ \bibinfo {pages}
  {023515} (\bibinfo {year} {2022})},\ \Eprint
  {http://arxiv.org/abs/2105.13544} {arXiv:2105.13544 [astro-ph.CO]}
  \BibitemShut {NoStop}%
\bibitem [{\citenamefont {Abbott}\ \emph {et~al.}(2023)\citenamefont {Abbott}
  \emph {et~al.}}]{Kilo-DegreeSurvey:2023gfr}%
  \BibitemOpen
  \bibfield  {author} {\bibinfo {author} {\bibfnamefont {T.~M.~C.}\
  \bibnamefont {Abbott}} \emph {et~al.} (\bibinfo {collaboration} {Kilo-Degree
  Survey, Dark Energy Survey}),\ }\href {\doibase 10.21105/astro.2305.17173}
  {\bibfield  {journal} {\bibinfo  {journal} {Open J. Astrophys.}\ }\textbf
  {\bibinfo {volume} {6}},\ \bibinfo {pages} {2305.17173} (\bibinfo {year}
  {2023})},\ \Eprint {http://arxiv.org/abs/2305.17173} {arXiv:2305.17173
  [astro-ph.CO]} \BibitemShut {NoStop}%
\bibitem [{\citenamefont {Rosenberg}\ \emph {et~al.}(2022)\citenamefont
  {Rosenberg}, \citenamefont {Gratton},\ and\ \citenamefont
  {Efstathiou}}]{Rosenberg:2022sdy}%
  \BibitemOpen
  \bibfield  {author} {\bibinfo {author} {\bibfnamefont {E.}~\bibnamefont
  {Rosenberg}}, \bibinfo {author} {\bibfnamefont {S.}~\bibnamefont {Gratton}},
  \ and\ \bibinfo {author} {\bibfnamefont {G.}~\bibnamefont {Efstathiou}},\
  }\href {\doibase 10.1093/mnras/stac2744} {\bibfield  {journal} {\bibinfo
  {journal} {Mon. Not. Roy. Astron. Soc.}\ }\textbf {\bibinfo {volume} {517}},\
  \bibinfo {pages} {4620} (\bibinfo {year} {2022})},\ \Eprint
  {http://arxiv.org/abs/2205.10869} {arXiv:2205.10869 [astro-ph.CO]}
  \BibitemShut {NoStop}%
\bibitem [{\citenamefont {Aghanim}\ \emph {et~al.}(2020)\citenamefont {Aghanim}
  \emph {et~al.}}]{Planck:2019nip}%
  \BibitemOpen
  \bibfield  {author} {\bibinfo {author} {\bibfnamefont {N.}~\bibnamefont
  {Aghanim}} \emph {et~al.} (\bibinfo {collaboration} {Planck}),\ }\href
  {\doibase 10.1051/0004-6361/201936386} {\bibfield  {journal} {\bibinfo
  {journal} {Astron. Astrophys.}\ }\textbf {\bibinfo {volume} {641}},\ \bibinfo
  {pages} {A5} (\bibinfo {year} {2020})},\ \Eprint
  {http://arxiv.org/abs/1907.12875} {arXiv:1907.12875 [astro-ph.CO]}
  \BibitemShut {NoStop}%
\bibitem [{\citenamefont {Carron}\ \emph {et~al.}(2022)\citenamefont {Carron},
  \citenamefont {Mirmelstein},\ and\ \citenamefont {Lewis}}]{Carron:2022eyg}%
  \BibitemOpen
  \bibfield  {author} {\bibinfo {author} {\bibfnamefont {J.}~\bibnamefont
  {Carron}}, \bibinfo {author} {\bibfnamefont {M.}~\bibnamefont {Mirmelstein}},
  \ and\ \bibinfo {author} {\bibfnamefont {A.}~\bibnamefont {Lewis}},\ }\href
  {\doibase 10.1088/1475-7516/2022/09/039} {\bibfield  {journal} {\bibinfo
  {journal} {JCAP}\ }\textbf {\bibinfo {volume} {09}},\ \bibinfo {pages} {039}
  (\bibinfo {year} {2022})},\ \Eprint {http://arxiv.org/abs/2206.07773}
  {arXiv:2206.07773 [astro-ph.CO]} \BibitemShut {NoStop}%
\bibitem [{\citenamefont {Garcia-Berro}\ \emph {et~al.}(1999)\citenamefont
  {Garcia-Berro}, \citenamefont {Gaztanaga}, \citenamefont {Isern},
  \citenamefont {Benvenuto},\ and\ \citenamefont
  {Althaus}}]{Garcia-Berro:1999cwy}%
  \BibitemOpen
  \bibfield  {author} {\bibinfo {author} {\bibfnamefont {E.}~\bibnamefont
  {Garcia-Berro}}, \bibinfo {author} {\bibfnamefont {E.}~\bibnamefont
  {Gaztanaga}}, \bibinfo {author} {\bibfnamefont {J.}~\bibnamefont {Isern}},
  \bibinfo {author} {\bibfnamefont {O.}~\bibnamefont {Benvenuto}}, \ and\
  \bibinfo {author} {\bibfnamefont {L.}~\bibnamefont {Althaus}},\ }\href@noop
  {} {\  (\bibinfo {year} {1999})},\ \Eprint
  {http://arxiv.org/abs/astro-ph/9907440} {arXiv:astro-ph/9907440} \BibitemShut
  {NoStop}%
\bibitem [{\citenamefont {Riazuelo}\ and\ \citenamefont
  {Uzan}(2002)}]{Riazuelo:2001mg}%
  \BibitemOpen
  \bibfield  {author} {\bibinfo {author} {\bibfnamefont {A.}~\bibnamefont
  {Riazuelo}}\ and\ \bibinfo {author} {\bibfnamefont {J.-P.}\ \bibnamefont
  {Uzan}},\ }\href {\doibase 10.1103/PhysRevD.66.023525} {\bibfield  {journal}
  {\bibinfo  {journal} {Phys. Rev. D}\ }\textbf {\bibinfo {volume} {66}},\
  \bibinfo {pages} {023525} (\bibinfo {year} {2002})},\ \Eprint
  {http://arxiv.org/abs/astro-ph/0107386} {arXiv:astro-ph/0107386} \BibitemShut
  {NoStop}%
\bibitem [{\citenamefont {Nesseris}\ and\ \citenamefont
  {Perivolaropoulos}(2006)}]{Nesseris:2006jc}%
  \BibitemOpen
  \bibfield  {author} {\bibinfo {author} {\bibfnamefont {S.}~\bibnamefont
  {Nesseris}}\ and\ \bibinfo {author} {\bibfnamefont {L.}~\bibnamefont
  {Perivolaropoulos}},\ }\href {\doibase 10.1103/PhysRevD.73.103511} {\bibfield
   {journal} {\bibinfo  {journal} {Phys. Rev. D}\ }\textbf {\bibinfo {volume}
  {73}},\ \bibinfo {pages} {103511} (\bibinfo {year} {2006})},\ \Eprint
  {http://arxiv.org/abs/astro-ph/0602053} {arXiv:astro-ph/0602053} \BibitemShut
  {NoStop}%
\bibitem [{\citenamefont {Wright}\ and\ \citenamefont
  {Li}(2018)}]{Wright:2017rsu}%
  \BibitemOpen
  \bibfield  {author} {\bibinfo {author} {\bibfnamefont {B.~S.}\ \bibnamefont
  {Wright}}\ and\ \bibinfo {author} {\bibfnamefont {B.}~\bibnamefont {Li}},\
  }\href {\doibase 10.1103/PhysRevD.97.083505} {\bibfield  {journal} {\bibinfo
  {journal} {Phys. Rev. D}\ }\textbf {\bibinfo {volume} {97}},\ \bibinfo
  {pages} {083505} (\bibinfo {year} {2018})},\ \Eprint
  {http://arxiv.org/abs/1710.07018} {arXiv:1710.07018 [astro-ph.CO]}
  \BibitemShut {NoStop}%
\bibitem [{\citenamefont {Will}(2014)}]{Will:2014kxa}%
  \BibitemOpen
  \bibfield  {author} {\bibinfo {author} {\bibfnamefont {C.~M.}\ \bibnamefont
  {Will}},\ }\href {\doibase 10.12942/lrr-2014-4} {\bibfield  {journal}
  {\bibinfo  {journal} {Living Rev. Rel.}\ }\textbf {\bibinfo {volume} {17}},\
  \bibinfo {pages} {4} (\bibinfo {year} {2014})},\ \Eprint
  {http://arxiv.org/abs/1403.7377} {arXiv:1403.7377 [gr-qc]} \BibitemShut
  {NoStop}%
\bibitem [{\citenamefont {Raveri}\ \emph {et~al.}(2023)\citenamefont {Raveri},
  \citenamefont {Pogosian}, \citenamefont {Martinelli}, \citenamefont {Koyama},
  \citenamefont {Silvestri},\ and\ \citenamefont {Zhao}}]{Raveri:2021dbu}%
  \BibitemOpen
  \bibfield  {author} {\bibinfo {author} {\bibfnamefont {M.}~\bibnamefont
  {Raveri}}, \bibinfo {author} {\bibfnamefont {L.}~\bibnamefont {Pogosian}},
  \bibinfo {author} {\bibfnamefont {M.}~\bibnamefont {Martinelli}}, \bibinfo
  {author} {\bibfnamefont {K.}~\bibnamefont {Koyama}}, \bibinfo {author}
  {\bibfnamefont {A.}~\bibnamefont {Silvestri}}, \ and\ \bibinfo {author}
  {\bibfnamefont {G.-B.}\ \bibnamefont {Zhao}},\ }\href {\doibase
  10.1088/1475-7516/2023/02/061} {\bibfield  {journal} {\bibinfo  {journal}
  {JCAP}\ }\textbf {\bibinfo {volume} {02}},\ \bibinfo {pages} {061} (\bibinfo
  {year} {2023})},\ \Eprint {http://arxiv.org/abs/2107.12990} {arXiv:2107.12990
  [astro-ph.CO]} \BibitemShut {NoStop}%
\bibitem [{\citenamefont {Pogosian}\ \emph {et~al.}(2022)\citenamefont
  {Pogosian}, \citenamefont {Raveri}, \citenamefont {Koyama}, \citenamefont
  {Martinelli}, \citenamefont {Silvestri}, \citenamefont {Zhao}, \citenamefont
  {Li}, \citenamefont {Peirone},\ and\ \citenamefont
  {Zucca}}]{Pogosian:2021mcs}%
  \BibitemOpen
  \bibfield  {author} {\bibinfo {author} {\bibfnamefont {L.}~\bibnamefont
  {Pogosian}}, \bibinfo {author} {\bibfnamefont {M.}~\bibnamefont {Raveri}},
  \bibinfo {author} {\bibfnamefont {K.}~\bibnamefont {Koyama}}, \bibinfo
  {author} {\bibfnamefont {M.}~\bibnamefont {Martinelli}}, \bibinfo {author}
  {\bibfnamefont {A.}~\bibnamefont {Silvestri}}, \bibinfo {author}
  {\bibfnamefont {G.-B.}\ \bibnamefont {Zhao}}, \bibinfo {author}
  {\bibfnamefont {J.}~\bibnamefont {Li}}, \bibinfo {author} {\bibfnamefont
  {S.}~\bibnamefont {Peirone}}, \ and\ \bibinfo {author} {\bibfnamefont
  {A.}~\bibnamefont {Zucca}},\ }\href {\doibase 10.1038/s41550-022-01808-7}
  {\bibfield  {journal} {\bibinfo  {journal} {Nature Astron.}\ }\textbf
  {\bibinfo {volume} {6}},\ \bibinfo {pages} {1484} (\bibinfo {year} {2022})},\
  \Eprint {http://arxiv.org/abs/2107.12992} {arXiv:2107.12992 [astro-ph.CO]}
  \BibitemShut {NoStop}%
\bibitem [{\citenamefont {Calderon}\ \emph {et~al.}(2024)\citenamefont
  {Calderon} \emph {et~al.}}]{DESI:2024aqx}%
  \BibitemOpen
  \bibfield  {author} {\bibinfo {author} {\bibfnamefont {R.}~\bibnamefont
  {Calderon}} \emph {et~al.} (\bibinfo {collaboration} {DESI}),\ }\href@noop {}
  {\  (\bibinfo {year} {2024})},\ \Eprint {http://arxiv.org/abs/2405.04216}
  {arXiv:2405.04216 [astro-ph.CO]} \BibitemShut {NoStop}%
\bibitem [{\citenamefont {Yang}\ \emph {et~al.}(2024)\citenamefont {Yang},
  \citenamefont {Ren}, \citenamefont {Wang}, \citenamefont {Lu}, \citenamefont
  {Zhang}, \citenamefont {Cai},\ and\ \citenamefont
  {Saridakis}}]{Yang:2024kdo}%
  \BibitemOpen
  \bibfield  {author} {\bibinfo {author} {\bibfnamefont {Y.}~\bibnamefont
  {Yang}}, \bibinfo {author} {\bibfnamefont {X.}~\bibnamefont {Ren}}, \bibinfo
  {author} {\bibfnamefont {Q.}~\bibnamefont {Wang}}, \bibinfo {author}
  {\bibfnamefont {Z.}~\bibnamefont {Lu}}, \bibinfo {author} {\bibfnamefont
  {D.}~\bibnamefont {Zhang}}, \bibinfo {author} {\bibfnamefont {Y.-F.}\
  \bibnamefont {Cai}}, \ and\ \bibinfo {author} {\bibfnamefont {E.~N.}\
  \bibnamefont {Saridakis}},\ }\href@noop {} {\  (\bibinfo {year} {2024})},\
  \Eprint {http://arxiv.org/abs/2404.19437} {arXiv:2404.19437 [astro-ph.CO]}
  \BibitemShut {NoStop}%
\bibitem [{\citenamefont {Hu}\ \emph {et~al.}(2014{\natexlab{a}})\citenamefont
  {Hu}, \citenamefont {Raveri}, \citenamefont {Frusciante},\ and\ \citenamefont
  {Silvestri}}]{Hu:2014oga}%
  \BibitemOpen
  \bibfield  {author} {\bibinfo {author} {\bibfnamefont {B.}~\bibnamefont
  {Hu}}, \bibinfo {author} {\bibfnamefont {M.}~\bibnamefont {Raveri}}, \bibinfo
  {author} {\bibfnamefont {N.}~\bibnamefont {Frusciante}}, \ and\ \bibinfo
  {author} {\bibfnamefont {A.}~\bibnamefont {Silvestri}},\ }\href@noop {} {\
  (\bibinfo {year} {2014}{\natexlab{a}})},\ \Eprint
  {http://arxiv.org/abs/1405.3590} {arXiv:1405.3590 [astro-ph.IM]} \BibitemShut
  {NoStop}%
\bibitem [{\citenamefont {Raveri}\ \emph {et~al.}(2017)\citenamefont {Raveri},
  \citenamefont {Bull}, \citenamefont {Silvestri},\ and\ \citenamefont
  {Pogosian}}]{Raveri:2017qvt}%
  \BibitemOpen
  \bibfield  {author} {\bibinfo {author} {\bibfnamefont {M.}~\bibnamefont
  {Raveri}}, \bibinfo {author} {\bibfnamefont {P.}~\bibnamefont {Bull}},
  \bibinfo {author} {\bibfnamefont {A.}~\bibnamefont {Silvestri}}, \ and\
  \bibinfo {author} {\bibfnamefont {L.}~\bibnamefont {Pogosian}},\ }\href
  {\doibase 10.1103/PhysRevD.96.083509} {\bibfield  {journal} {\bibinfo
  {journal} {Phys. Rev. D}\ }\textbf {\bibinfo {volume} {96}},\ \bibinfo
  {pages} {083509} (\bibinfo {year} {2017})},\ \Eprint
  {http://arxiv.org/abs/1703.05297} {arXiv:1703.05297 [astro-ph.CO]}
  \BibitemShut {NoStop}%
\bibitem [{\citenamefont {Wolf}\ and\ \citenamefont
  {Ferreira}(2023)}]{Wolf:2023uno}%
  \BibitemOpen
  \bibfield  {author} {\bibinfo {author} {\bibfnamefont {W.~J.}\ \bibnamefont
  {Wolf}}\ and\ \bibinfo {author} {\bibfnamefont {P.~G.}\ \bibnamefont
  {Ferreira}},\ }\href {\doibase 10.1103/PhysRevD.108.103519} {\bibfield
  {journal} {\bibinfo  {journal} {Phys. Rev. D}\ }\textbf {\bibinfo {volume}
  {108}},\ \bibinfo {pages} {103519} (\bibinfo {year} {2023})},\ \Eprint
  {http://arxiv.org/abs/2310.07482} {arXiv:2310.07482 [astro-ph.CO]}
  \BibitemShut {NoStop}%
\bibitem [{\citenamefont {Hu}\ and\ \citenamefont {Sawicki}(2007)}]{Hu:2007pj}%
  \BibitemOpen
  \bibfield  {author} {\bibinfo {author} {\bibfnamefont {W.}~\bibnamefont
  {Hu}}\ and\ \bibinfo {author} {\bibfnamefont {I.}~\bibnamefont {Sawicki}},\
  }\href {\doibase 10.1103/PhysRevD.76.104043} {\bibfield  {journal} {\bibinfo
  {journal} {Phys. Rev. D}\ }\textbf {\bibinfo {volume} {76}},\ \bibinfo
  {pages} {104043} (\bibinfo {year} {2007})},\ \Eprint
  {http://arxiv.org/abs/0708.1190} {arXiv:0708.1190 [astro-ph]} \BibitemShut
  {NoStop}%
\bibitem [{\citenamefont {Hu}\ \emph {et~al.}(2014{\natexlab{b}})\citenamefont
  {Hu}, \citenamefont {Raveri}, \citenamefont {Frusciante},\ and\ \citenamefont
  {Silvestri}}]{Hu:2013twa}%
  \BibitemOpen
  \bibfield  {author} {\bibinfo {author} {\bibfnamefont {B.}~\bibnamefont
  {Hu}}, \bibinfo {author} {\bibfnamefont {M.}~\bibnamefont {Raveri}}, \bibinfo
  {author} {\bibfnamefont {N.}~\bibnamefont {Frusciante}}, \ and\ \bibinfo
  {author} {\bibfnamefont {A.}~\bibnamefont {Silvestri}},\ }\href {\doibase
  10.1103/PhysRevD.89.103530} {\bibfield  {journal} {\bibinfo  {journal} {Phys.
  Rev. D}\ }\textbf {\bibinfo {volume} {89}},\ \bibinfo {pages} {103530}
  (\bibinfo {year} {2014}{\natexlab{b}})},\ \Eprint
  {http://arxiv.org/abs/1312.5742} {arXiv:1312.5742 [astro-ph.CO]} \BibitemShut
  {NoStop}%
\bibitem [{\citenamefont {Raveri}\ \emph {et~al.}(2014)\citenamefont {Raveri},
  \citenamefont {Hu}, \citenamefont {Frusciante},\ and\ \citenamefont
  {Silvestri}}]{Raveri:2014cka}%
  \BibitemOpen
  \bibfield  {author} {\bibinfo {author} {\bibfnamefont {M.}~\bibnamefont
  {Raveri}}, \bibinfo {author} {\bibfnamefont {B.}~\bibnamefont {Hu}}, \bibinfo
  {author} {\bibfnamefont {N.}~\bibnamefont {Frusciante}}, \ and\ \bibinfo
  {author} {\bibfnamefont {A.}~\bibnamefont {Silvestri}},\ }\href {\doibase
  10.1103/PhysRevD.90.043513} {\bibfield  {journal} {\bibinfo  {journal} {Phys.
  Rev. D}\ }\textbf {\bibinfo {volume} {90}},\ \bibinfo {pages} {043513}
  (\bibinfo {year} {2014})},\ \Eprint {http://arxiv.org/abs/1405.1022}
  {arXiv:1405.1022 [astro-ph.CO]} \BibitemShut {NoStop}%
\bibitem [{\citenamefont {Lewis}\ \emph {et~al.}(2000)\citenamefont {Lewis},
  \citenamefont {Challinor},\ and\ \citenamefont {Lasenby}}]{Lewis:1999bs}%
  \BibitemOpen
  \bibfield  {author} {\bibinfo {author} {\bibfnamefont {A.}~\bibnamefont
  {Lewis}}, \bibinfo {author} {\bibfnamefont {A.}~\bibnamefont {Challinor}}, \
  and\ \bibinfo {author} {\bibfnamefont {A.}~\bibnamefont {Lasenby}},\ }\href
  {\doibase 10.1086/309179} {\bibfield  {journal} {\bibinfo  {journal}
  {Astrophys. J.}\ }\textbf {\bibinfo {volume} {538}},\ \bibinfo {pages} {473}
  (\bibinfo {year} {2000})},\ \Eprint {http://arxiv.org/abs/astro-ph/9911177}
  {arXiv:astro-ph/9911177 [astro-ph]} \BibitemShut {NoStop}%
%%CITATION = ASTRO-PH/9911177;%%
\bibitem [{\citenamefont {Torrado}\ and\ \citenamefont
  {Lewis}(2021)}]{Torrado:2020dgo}%
  \BibitemOpen
  \bibfield  {author} {\bibinfo {author} {\bibfnamefont {J.}~\bibnamefont
  {Torrado}}\ and\ \bibinfo {author} {\bibfnamefont {A.}~\bibnamefont
  {Lewis}},\ }\href {\doibase 10.1088/1475-7516/2021/05/057} {\bibfield
  {journal} {\bibinfo  {journal} {JCAP}\ }\textbf {\bibinfo {volume} {05}},\
  \bibinfo {pages} {057} (\bibinfo {year} {2021})},\ \Eprint
  {http://arxiv.org/abs/2005.05290} {arXiv:2005.05290 [astro-ph.IM]}
  \BibitemShut {NoStop}%
\bibitem [{\citenamefont {{Torrado}}\ and\ \citenamefont
  {{Lewis}}(2019)}]{2019ascl.soft10019T}%
  \BibitemOpen
  \bibfield  {author} {\bibinfo {author} {\bibfnamefont {J.}~\bibnamefont
  {{Torrado}}}\ and\ \bibinfo {author} {\bibfnamefont {A.}~\bibnamefont
  {{Lewis}}},\ }\href@noop {} {\enquote {\bibinfo {title} {{Cobaya: Bayesian
  analysis in cosmology}},}\ }\bibinfo {howpublished} {Astrophysics Source Code
  Library, record ascl:1910.019} (\bibinfo {year} {2019})\BibitemShut {NoStop}%
\bibitem [{\citenamefont {Lewis}(2019)}]{Lewis:2019xzd}%
  \BibitemOpen
  \bibfield  {author} {\bibinfo {author} {\bibfnamefont {A.}~\bibnamefont
  {Lewis}},\ }\href {https://getdist.readthedocs.io} {\  (\bibinfo {year}
  {2019})},\ \Eprint {http://arxiv.org/abs/1910.13970} {arXiv:1910.13970
  [astro-ph.IM]} \BibitemShut {NoStop}%
%%CITATION = ARXIV:1910.13970;%%
\end{thebibliography}%

\end{document}